\documentclass[a4paper, 11pt]{article}
\usepackage{jinsttemplate}

\usepackage[colorlinks=true,linkcolor=blue,citecolor=blue,urlcolor=blue]{hyperref}
\usepackage{graphicx}
\usepackage{float}
\usepackage{gensymb}
\usepackage{makecell}
\usepackage{array}
\usepackage{booktabs}
\usepackage{url}
\usepackage[skip=10pt]{caption}
\usepackage{subcaption}

\graphicspath{{figures/}}

\usepackage[mathlines]{lineno}

\usepackage{verbatim}

\title{Long-term scintillation light performance, calibrations, and triggering efficiency in the MicroBooNE detector}
\collaboration{MicroBooNE Collaboration}

\begin{document}

\author[mm]{P.~Abratenko}
\author[n]{D.~Andrade~Aldana}
\author[v]{L.~Arellano}
\author[ll]{J.~Asaadi}
\author[kk]{A.~Ashkenazi}
\author[l]{S.~Balasubramanian}
\author[l]{B.~Baller}
\author[dd]{A.~Barnard}
\author[dd]{G.~Barr}
\author[dd]{D.~Barrow}
% ...

% ...
\affiliation[l]{Fermi National Accelerator Laboratory (FNAL), Batavia, IL 60510, USA}
% ...
\affiliation[n]{Illinois Institute of Technology (IIT), Chicago, IL 60616, USA}
% ...
\affiliation[v]{The University of Manchester, Manchester M13 9PL, United Kingdom}
% ...
\affiliation[dd]{University of Oxford, Oxford OX1 3RH, United Kingdom}
% ...
\affiliation[kk]{Tel Aviv University, Tel Aviv, Israel, 69978}
\affiliation[ll]{University of Texas, Arlington, TX, 76019, USA}
\affiliation[mm]{Tufts University, Medford, MA, 02155, USA}
% ...

\emailAdd{microboone\_info@fnal.gov}
\date{\today}

\abstract{A detailed understanding of scintillation light in liquid argon time projection chambers (LArTPCs) is essential for the robust extraction of physics measurements from their data. This article presents the long-term performance of scintillation light production and detection using five years of data from the MicroBooNE LArTPC neutrino detector. We describe the MicroBooNE light simulation and light-based triggering schemes, and present the photomultiplier tube gain calibration and stability, light response stability, and light-based systematic uncertainties over the course of five years of data collection. We also make an in-situ measurement of the scintillation light triggering efficiency down to 30 MeV of visible energy. Finally, we discuss two notable observations in MicroBooNE's data, both reported here for the first time: an approximately 50\% decline in MicroBooNE's light yield over time, concentrated in the first two years of running; and a higher than expected $\mathcal{O}$(200~kHz) rate of single photoelectrons (SPE). We present the first measurements of the dependence of the SPE rate on the TPC electric field strength and polarity. Together, the results presented in this article provide an important benchmark for the long-term performance of scintillation light systems in LArTPCs and their use in particle physics measurements.
}


\maketitle
\section{Introduction}

MicroBooNE is a Liquid Argon Time Projection Chamber (LArTPC) neutrino detector located along the Booster Neutrino Beam (BNB) and off-axis to the Neutrinos at the Main Injector (NuMI) beam at the Fermi National Accelerator Laboratory. MicroBooNE operated from 2015-2021, accumulating 5 years of physics data and several post-data-taking R\&D runs. MicroBooNE's primary physics goal is to investigate possible explanations for the short-baseline low energy excess observed by LSND~\cite{LSND:2001aii} and MiniBooNE~\cite{MiniBooNE:2020pnu} through various avenues~\cite{PhysRevLett.128.241801, MicroBooNE:2025nll}. Alongside this, MicroBooNE has broad programs in neutrino interaction physics, beyond the Standard Model physics, and detector R\&D for next-generation LArTPCs. 

LArTPC detectors combine precision tracking of charged particles with calorimetry, enabling excellent particle identification and energy reconstruction. Interacting neutrinos produce charged particles that ionise the liquid argon. The ionisation charge is drifted, by an applied electric field, towards segmented charge readout wire planes, where it is collected. In addition, liquid argon is a prolific scintillator: in the absence of electric field, approximately 40,000 vacuum-ultraviolet (VUV) scintillation light photons with wavelength of 128~nm are emitted per MeV of energy loss~\cite{Doke:1990rza}. The photons are emitted from the de-excitation of excited argon dimers with two distinct time constants: a singlet (prompt) state with $\tau_s \sim 6$\,ns and a triplet (slow) state with $\tau_t \sim 1.3-1.6$\,$\upmu$s~\cite{Hitachi:1983zz, Segreto:2014aia}. The scintillation light propagates through the detector much faster than the ionisation charge, providing precise timing information. This is used to determine the interaction time, associate charge deposits with the interaction, and distinguish activity in time with the neutrino beam from non-beam backgrounds.

In MicroBooNE, scintillation light provides the primary trigger for recording beam neutrino interactions and enables the suppression of cosmic-ray-induced backgrounds. The efficiency of these selections depends directly on the amount of light detected, and therefore on the stability and monitoring of the optical response. Changes in photomultiplier tube (PMT) gain, scintillation-light yield and light propagation can alter event-selection efficiencies, particularly when close to trigger thresholds, and must either be calibrated or incorporated as systematic uncertainties. Understanding these effects is particularly important for low-energy and rare-event searches, and for future LArTPC experiments expected to operate over many years including the Short-Baseline Neutrino Program~\cite{Machado:2019oxb} and the Deep Underground Neutrino Experiment~\cite{DUNE:2020lwj}. 

MicroBooNE’s five years of operation provide an opportunity to study these effects over one of the longest running periods of a large LArTPC neutrino detector. In this article, we present the long-term performance of scintillation light production and detection in MicroBooNE. We describe the MicroBooNE light simulation, light-based triggering schemes, photomultiplier tube gain calibration and stability, light response stability, and light-based systematic uncertainties over the course of five years of data collection. We also present an in-situ measurement of the scintillation light triggering efficiency down to 30 MeV of visible energy. Finally, we discuss two notable observations in MicroBooNE's data, both reported here for the first time: an approximately 50\% decline in MicroBooNE's light yield over time, concentrated in the first two years of running; and a higher than expected $\mathcal{O}$(200~kHz) rate of single photoelectrons (SPE). We discuss investigations into the origin of the light yield decline, including testing samples of argon for contaminants, and we make the first measurements of the dependence of the SPE rate on the TPC electric field strength and polarity. Together, the results presented in this article provide a benchmark for the long-term performance of scintillation light systems in LArTPCs and their use in particle physics measurements.

\section{Detector, simulation and triggering}

This section describes the MicroBooNE detector, light detection system, light simulation, and light-based triggering. 

\subsection{The MicroBooNE detector}

The MicroBooNE detector consists of a rectangular time projection chamber (TPC) housed inside a cylindrical cryostat containing 170~tonnes of liquid argon. The TPC has dimensions of 2.6\,m in the drift direction (horizontal, X), 2.3\,m vertically (Y), and 10.4\,m along the beam direction (Z), with a total instrumented liquid argon mass of 85~tonnes. An electric field of 273\,V/cm is applied across the TPC which drifts ionisation charge towards three planes of sense-wires with a maximum drift time of 2.3~ms. These planes are oriented vertically and at $\pm60\degree$ to the vertical with wire-spacings of 3\,mm.

Layers of Cosmic Ray Taggers (CRTs) are placed on the top and along the sides of the cryostat to primarily identify non–beam-related events. The CRT became fully operational in March 2017, part way through MicroBooNE's operations. Each CRT panel is composed of plastic scintillator strips with embedded wavelength-shifting fibres read out by silicon photomultipliers, and produces time-stamped scintillation signals when a charged particle traverses it. These time-stamped signals (“CRT hits”) provide an external position and timing tag for cosmic-ray muons, enabling the matching of CRT information to TPC tracks used throughout this article. A detailed description of the MicroBooNE detector and its components is found in ref.~\cite{MicroBooNE:2016pwy, MicroBooNE:2019lta}.

\subsection{Light detection system}

\begin{figure}[ht]
    \centering
    \begin{subfigure}{0.49\textwidth}
        \centering
        \includegraphics[width=\linewidth]{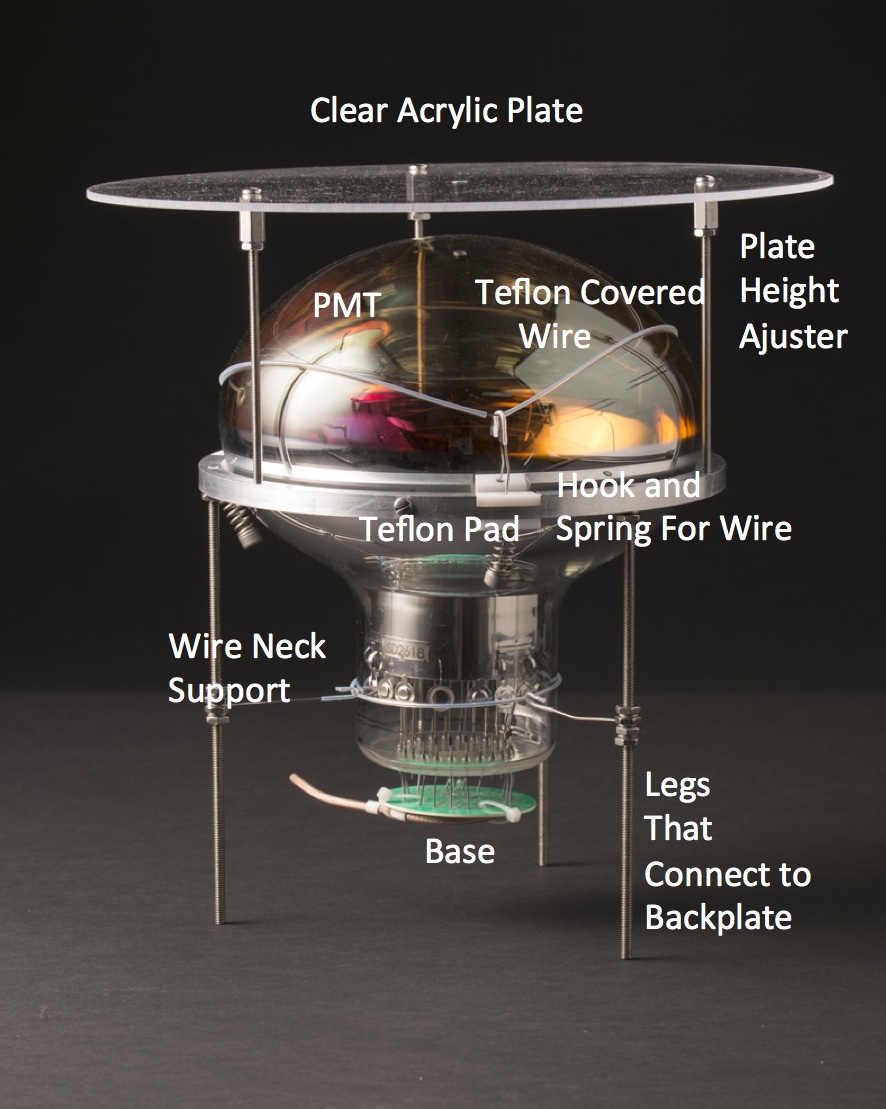}
        \caption{}
        \label{fig:pmt_photo_a}
    \end{subfigure}
    \begin{subfigure}{0.49\textwidth}
        \centering
        \includegraphics[width=\linewidth]{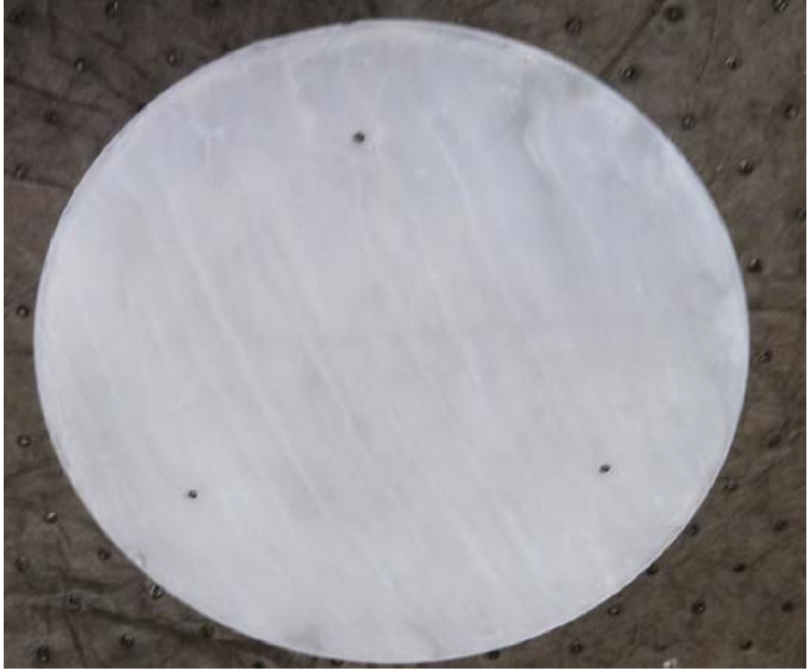}
        \caption{}
        \label{fig:pmt_photo_b}
    \end{subfigure}
    \caption{(a) Photograph of a MicroBooNE PMT optical unit. (b) Photograph of the TPB-coated acrylic plate. Photographs taken from~\cite{MicroBooNE:2016pwy}.}
    \label{fig:pmt_photo}
\end{figure}

\begin{figure}
\centering
\includegraphics[width=0.99\textwidth]{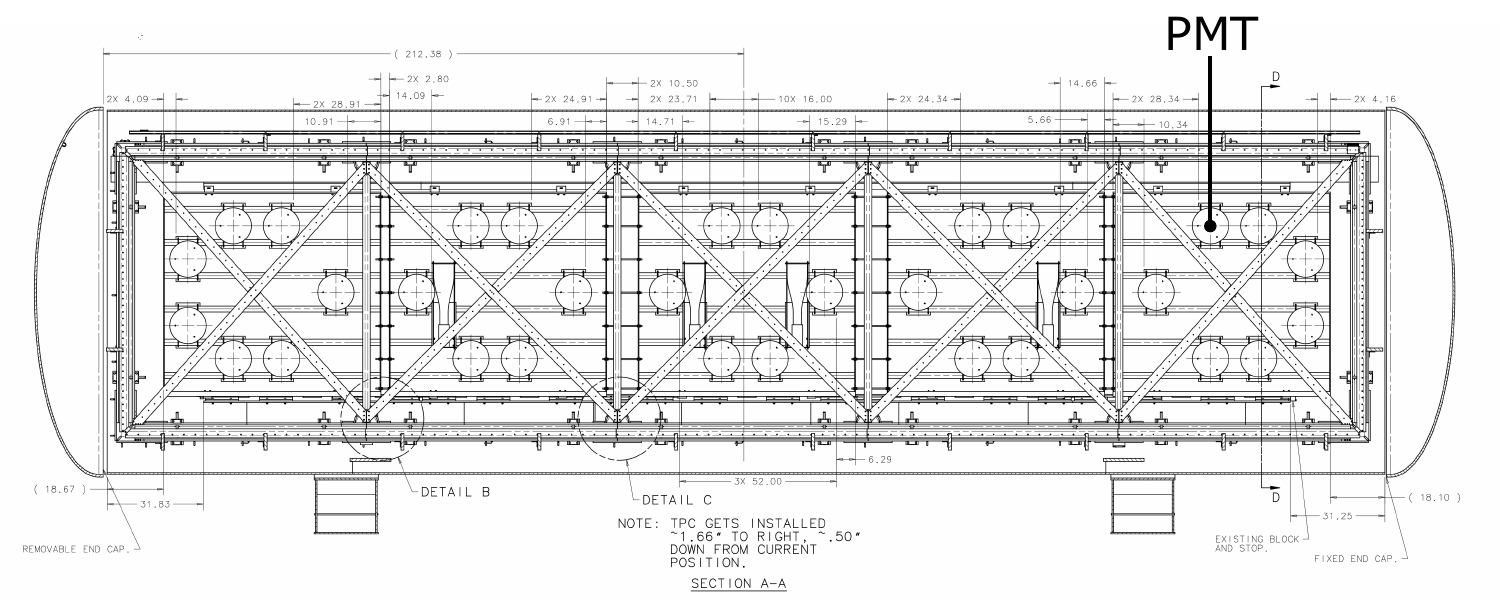}
\caption{Schematic of a cutout of the MicroBooNE TPC showing the placement of the light collection system. Each optical unit consists of an 8" PMT with a TPB-coated acrylic plate mounted in front. Figure adapted from~\cite{MicroBooNE:2016pwy}.}
\label{fig:tpc_light}
\end{figure}

The MicroBooNE primary light detection system consists of an array of 32 Hamamatsu R912-02mod cryogenic photomultiplier tubes (PMTs), each with a diameter of 8 inches, shown in figure~\ref{fig:pmt_photo}, located behind the sense-wire planes. Details of the PMTs used in MicroBooNE are found in ref.~\cite{Briese:2013wua}. The PMTs are arranged in 5 rosettes spanning the length of the detector, as shown in figure~\ref{fig:tpc_light}. In front of each PMT is a circular acrylic plate that is coated with a layer of tetraphenyl butadiene (TPB), a wavelength shifter. The TPB layer absorbs incident VUV light emitted by the liquid argon, and promptly re-emits visible light, peaking at a re-emission wavelength of 430~nm, where the PMTs have a high quantum efficiency. The PMTs are operated at a nominal high voltage (HV) of around 1,300\,V resulting in an operational electronic gain of $\sim10^7$. Each time the PMT system is turned off, the gains are recalibrated by performing minor adjustments to the high voltage. Additionally, MicroBooNE has a secondary light detection system made up of 4 light guide paddles~\cite{MicroBooNE:2016pwy}. This secondary system is not used in analysis and is not discussed in this article.  

For every triggered event, 23.4\,$\upmu$s of data is recorded from each PMTs. In dedicated “splitter boards”, the analogue signals are separated from the HV line via coupling capacitors. To ensure most waveforms are fully contained within the dynamic range of the digitizers, the signals are scaled down. Two copies of each waveform are ultimately saved: a high-gain (scaled to 18\% of the original amplitude) and a low-gain (scaled to 1.8\%) channel. The purpose of having both low-gain and high-gain channels is to increase the dynamic range of recorded data. Next, the waveforms pass through a pre-amplifier and a shaper with characteristic time of 60\,ns. Finally, they are digitized at 64\,MHz. High-gain waveforms are used for trigger formation and serve as the primary signal readout, whereas low-gain waveforms provide complementary information in events where the high-gain channels saturate.

\subsection{Light simulation}{\label{sec:light_sim}}

The large amount of scintillation light production in liquid argon makes simulation of photon propagation through the detector computationally challenging. Instead of tracing the path of each photon individually for each simulated event, pre-generated \textit{photon lookup libraries} are used to model the light production and propagation. To create these libraries, the whole cryostat volume is divided into 75 (X coordinate) × 75 (Y coordinate) × 400 (Z coordinate) voxels. Then, for each voxel a large number of photons are generated isotropically and their propagation to the photon detectors is simulated using Geant4~\cite{AGOSTINELLI2003250}, taking into account propagation effects including scattering, attenuation, and the reflectivities of all the materials in the active volume of the detector. The properties of the liquid argon used in the simulation are summarised in table~\ref{tab:light_argon_properties}. The photon transmission through the three wire planes prior to reaching the PMTs is corrected for geometrically based on the photon's incident angle. The result of the simulation is then convolved with the expected electric field map of the detector, both inside and outside the field cage, to account for the anti-correlation between light and charge due to the electric field and recombination effects. The convolved acceptance of each PMT for each voxel is evaluated and stored in the photon library as a map of probabilities, called \textit{visibilities}. In this context, the visibility can be understood as the fraction of generated photons recorded by the PMT.

\begin{table}[ht]
    \caption{Physical parameters used in MicroBooNE's light simulation taken from~\cite{MicroBooNE:2016pwy,ISHIDA1997380, Hitachi:1983zz, PhysRevB.17.2762}.}
    \centering
    \begin{tabular}{c c}
        \hline
         Physical property & Default value \\\hline \hline
        Absorption length & 20 m \\
        Rayleigh scattering length (@ 128 nm) & 66 cm  \\
        Singlet, Triplet state time constants & 6 ns, 1.6 $\upmu$s \\ 
        Photons/MeV for E = 0 V/cm & 40,000 \\
        Effective quantum efficiency (including TPB re-emission) & 0.93\% \\ \hline
    \end{tabular}
    
    \label{tab:light_argon_properties}
\end{table}

\begin{figure}[h!]
    \centering
    \begin{subfigure}{0.40\textwidth}
        \centering
        \includegraphics[width=\linewidth]{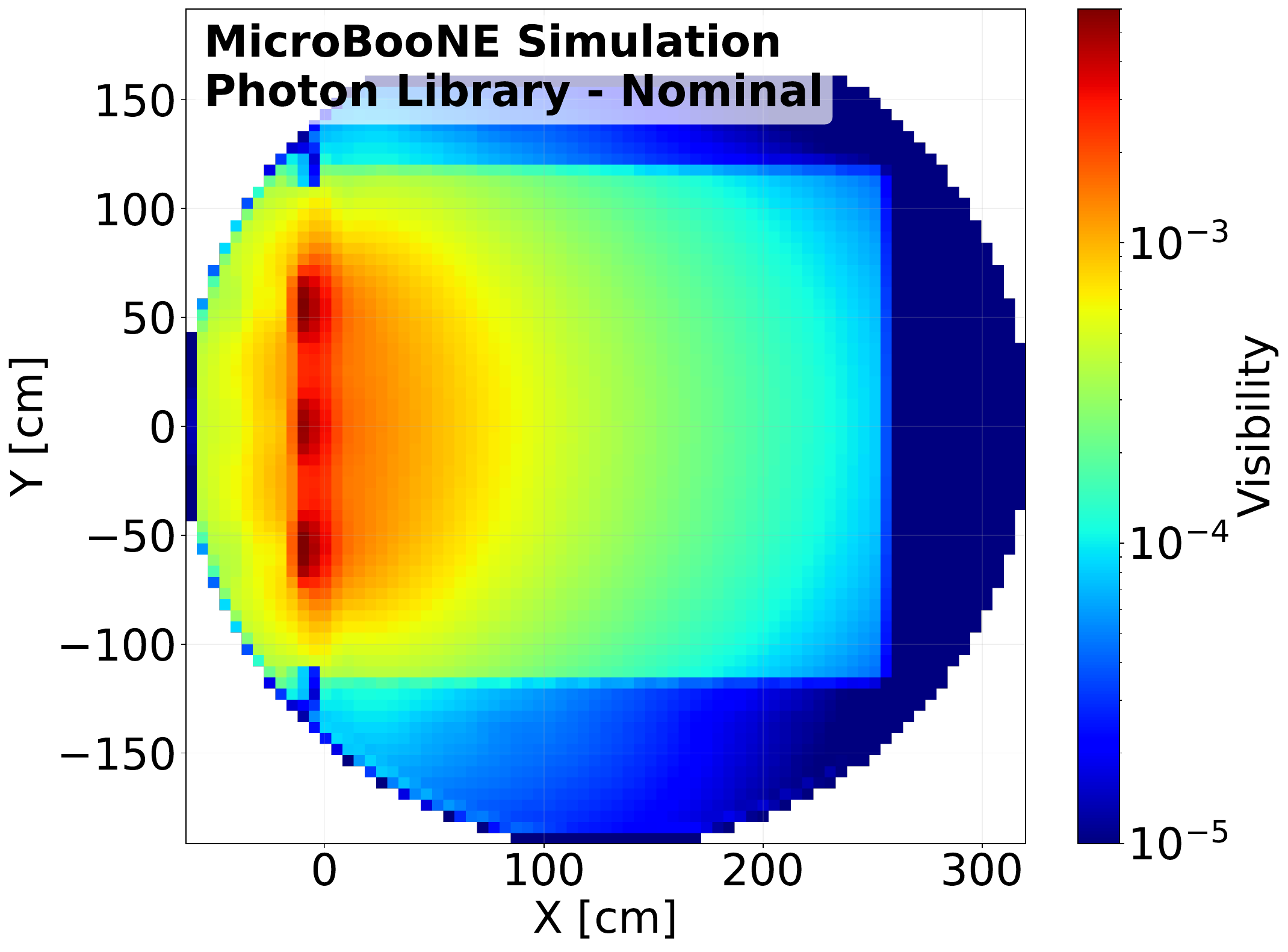}
        \caption{}
        \label{fig:cv_library_a}
    \end{subfigure}
    \hfill
    \begin{subfigure}{0.59\textwidth}
        \centering
        \includegraphics[width=\linewidth]{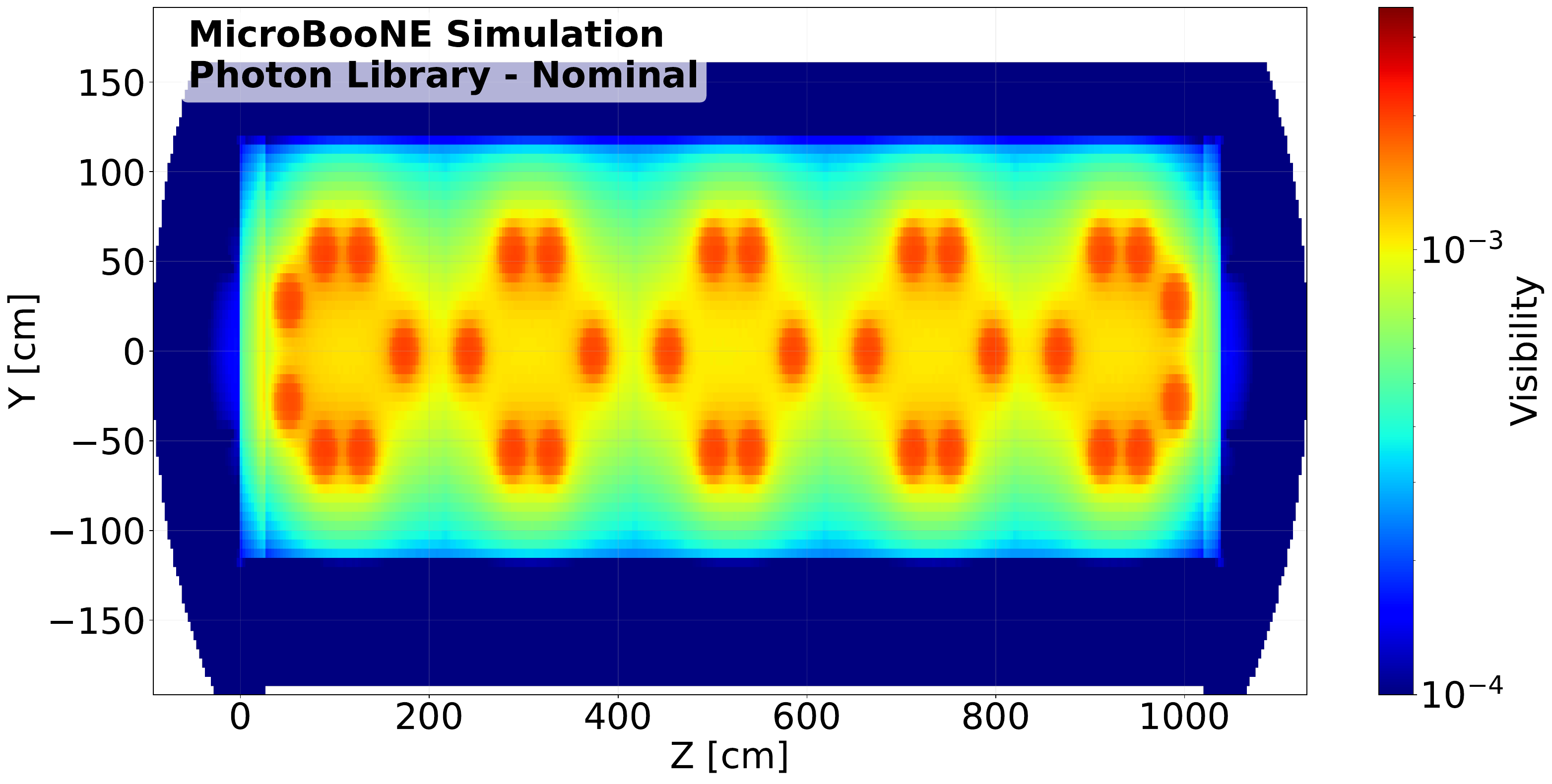}
        \caption{}
        \label{fig:cv_library_b}
    \end{subfigure}
    \caption{(a) Sliced view in the XY plane and (b) sliced view in the YZ plane of MicroBooNE's standard photon visibility library (expected fraction of photons accepted) used in photon simulation. The cylindrical shape of the cryostat and the rectangular TPC volume are visible through the borders of the photon visibility in both projections.}
    \label{fig:cv_library}
\end{figure}

Figure~\ref{fig:cv_library} presents the standard photon library employed in MicroBooNE, displayed as sliced projections in the YX and YZ planes. The geometry reveals the cylindrical cryostat enclosing the rectangular TPC volume bounded by the field cage. Visibility is set to zero outside the cryostat, consistent with the assumption of an optically opaque boundary. For the YX plane slice it is evident that photons generated closer to the PMTs (located at $x \simeq 0$\,cm) have higher visibility. The effect of the opaque cathode is shown with the cut off at $x \simeq 256$\,cm. The YZ plane slice shows the rosette patterns of the PMTs with the circular acceptance window of the acrylic plate.

In addition to the scintillation light, Cherenkov light is emitted as particles traverse the liquid argon. However, the Cherenkov light emission is subdominant relative to the scintillation light, producing only 1-2\% of the total emitted photons~\cite{Antonello:2004sx}. For computational efficiency, the Cherenkov light is not included in the MicroBooNE simulation. Given the small amount of Cherenkov light emission, its omission from the simulation is not expected to impact MicroBooNE physics measurements. 

\subsection{Light signals and triggers}

The amount of data that can be saved by the data acquisition system (DAQ) is limited by the transfer bandwidth and available disk storage. It is therefore challenging to record an event for every beam spill. Beam neutrinos interact in MicroBooNE in approximately 1/1000 of BNB beam spills and 1/50 NuMI beam spills~\cite{MicroBooNE:2016pwy}. To identify events of interest, MicroBooNE uses multiple hardware and software trigger chains that are applied both online (in real time) and offline during data processing. 

\begin{figure}[]
 \centering
 \includegraphics[width=1.0\textwidth]{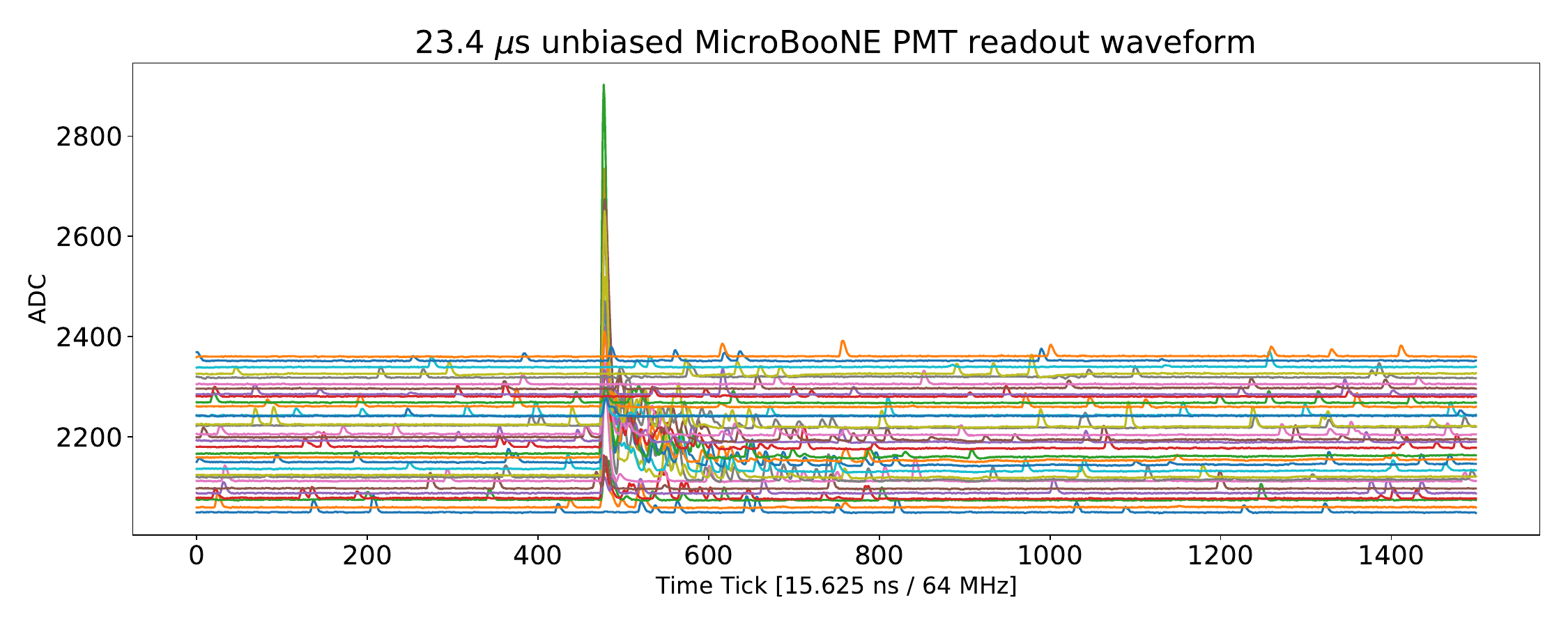}
 \caption{Example of a beam neutrino event producing light in the TPC with the waveforms recorded by the MicroBooNE PMTs. Each line represents one of the 32 PMT waveforms, aligned to the reconstructed interaction time. The baseline ADC values were adjusted for visualisation to avoid crowding.}
  \label{fig:beam_waveforms}
\end{figure}

The PMTs are used to trigger on beam neutrinos interacting in the TPC and locate the interaction vertex. When a neutrino interacts, a subset of the PMTs will collect large light signals as the photons from the interaction reach them. This produces an in-time coincidence spike in the PMT waveforms as shown in figure~\ref{fig:beam_waveforms} where more than half of the PMTs collected a large amount of light in analogue-to-digital converter (ADC) units at the same time. The large initial peak is caused by the prompt singlet state of the excited argon dimers while the tail is caused by the longer triplet state.

For both the BNB and NuMI beams, accelerator timing signals indicate when neutrinos are expected to arrive at the MicroBooNE detector. These timing signals generate the hardware trigger, which initiates data acquisition by instructing the detector to read out a complete TPC beam-gate window spanning 1.6~ms before and 3.2~ms after the trigger time. This extended window accounts for the comparatively slow drift velocity of electrons in liquid argon. 

The same beam timing signals are also distributed to the PMT readout electronics. Each PMT readout board receives dedicated beam-gate markers that arrive a few microseconds before the spill: approximately 4~µs early for the BNB and 10~µs for the NuMI beam. These early markers instruct the front-end electronics to begin continuous, uncompressed digitization of PMT waveforms over a 23.4~µs period encompassing the beam spill, ensuring that all prompt and delayed scintillation light activity is recorded. For each beam event, four 1.6~ms PMT data frames (one preceding, one containing, and two following the trigger) are stored. Outside of these beam windows, zero-suppression is applied to retain only regions of interest above threshold, thereby reducing data volume.

\begin{figure}[]
 \centering
 \includegraphics[width=0.8\textwidth]{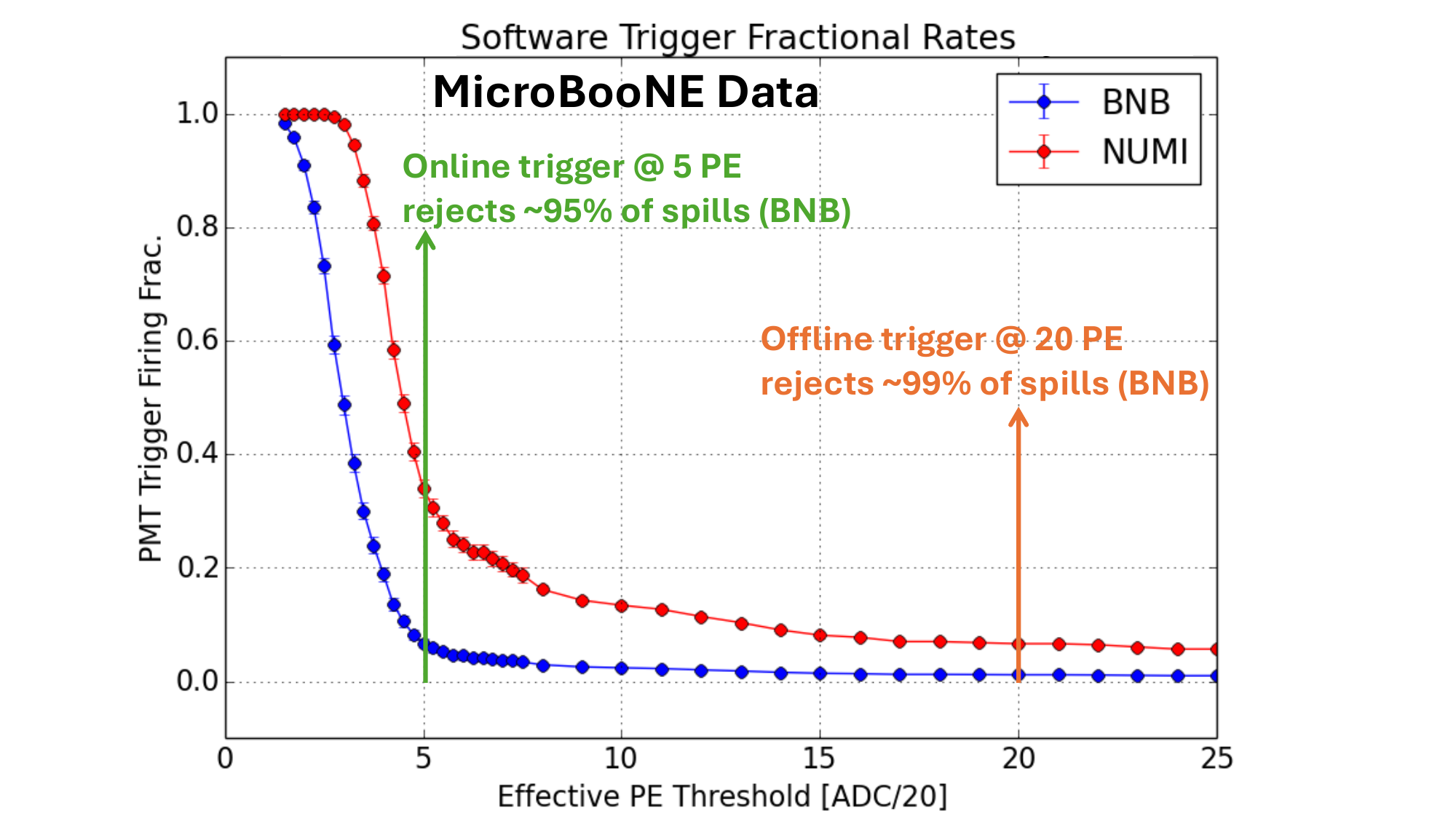}
 \caption{Software trigger acceptance rate for the BNB (blue) and NuMI (red), using a time window matched to each beam’s respective spill width. The rates were measured in off-beam mode, where cosmic-ray events dominate even with the software trigger applied.}
  \label{fig:trig_rate}
 \end{figure}

The first stage of the software trigger chain, the online trigger, requires the total number of reconstructed photoelectrons (PE) measured by the PMTs, $\Sigma_\mathrm{PE}$, to exceed $\sim$5~PE ($\sim$100~ADC) amplitude in any 100~ns time window during the BNB beam-spill gate. The $\Sigma_\mathrm{PE}$ is calculated by summing the PE from the 32 PMTs that each record at least 0.5~PE above baseline. Events satisfying this condition are tagged as beam interaction candidates, reducing empty events by about 95\% for the BNB. Figure~\ref{fig:trig_rate} shows the software-trigger acceptance rate as a function of $\Sigma_\mathrm{PE}$ for the BNB and NuMI beams. Both distributions exhibit a sharp drop after a few PE due to suppression of high-rate single-PE, followed by a more gradual decline as cosmic rays become the dominant light source. A consistently higher acceptance rate is seen for the NuMI beam than the BNB due to the 6 times longer NuMI beam spill.

Events passing the online trigger are recorded and stored for offline analysis, after which the offline (software) trigger determines whether they proceed to reconstruction and analysis. Using the same logic as the online trigger but with a higher threshold of 20~PE, this offline trigger is optimised to reduce data volume while preserving efficiency for low-energy neutrino interactions, rejecting approximately 99\% of empty beam spills, as shown in figure~\ref{fig:trig_rate}. The offline trigger was introduced part way through detector operations in 2016. The 20 PE threshold was chosen to keep the data processing requirements manageable, while maintaining a high efficiency BNB $\nu_\mu$ charged-current interactions with neutrino energies of 200--600~MeV. Events with between $\sim$5 and 20~PE remain in the recorded dataset and can be reprocessed with a lower offline threshold for future low-energy analyses.

\begin{figure}[]
 \centering
 \includegraphics[width=0.8\textwidth]{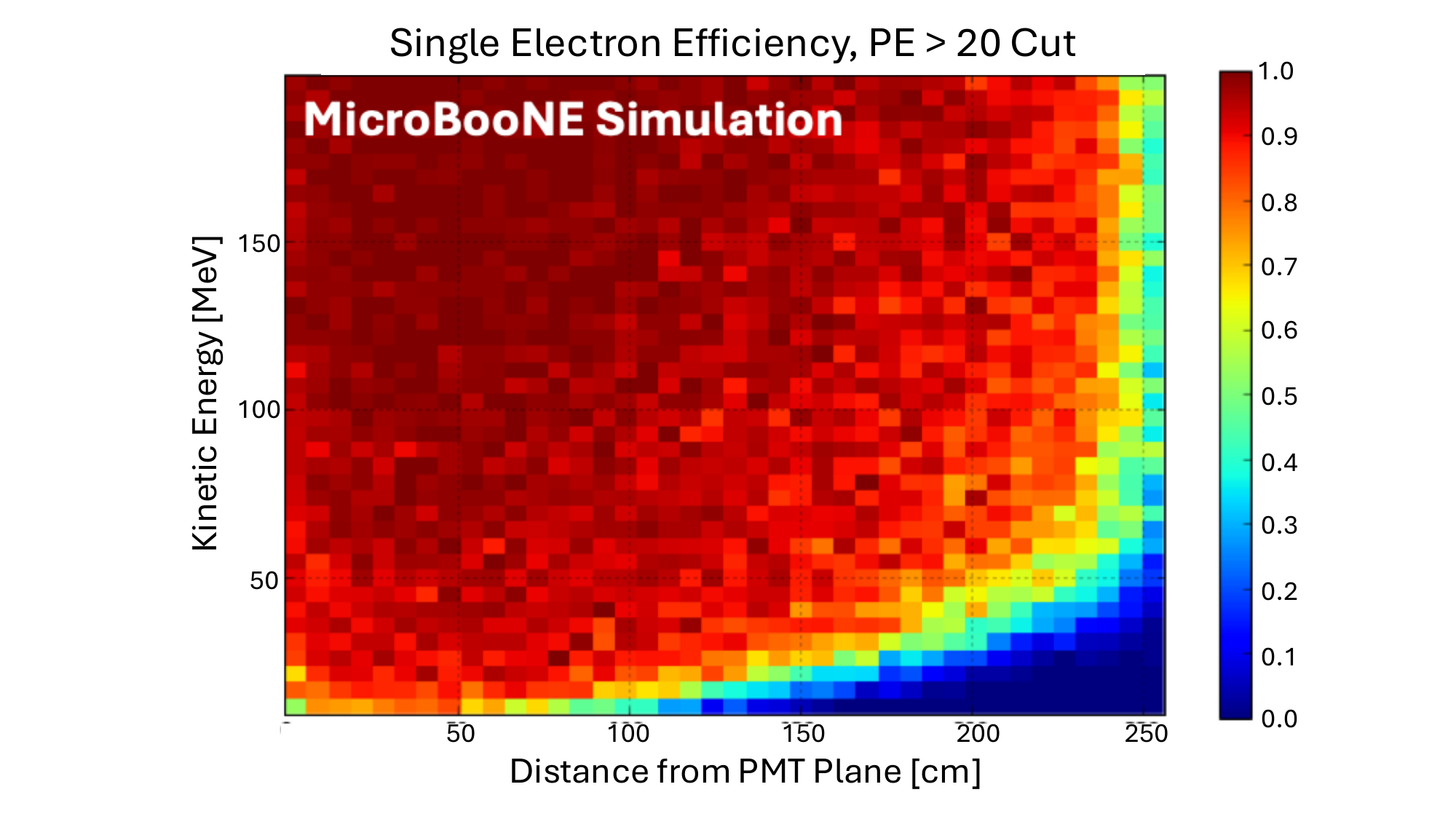}
 \caption{Fraction of single electron simulated events that pass the software trigger with a threshold of 20 PE. The events are shown with kinetic energy of the electrons and their distance from the PMT plane.}
  \label{fig:sim_trig}
 \end{figure}

To quantify the effect of this threshold, the software trigger efficiency was evaluated using simulated single-electron events. Figure~\ref{fig:sim_trig} shows the efficiency as a function of electron kinetic energy and distance from the PMT plane. The efficiency is expected to be very high but significantly decreases for lower energy events and events located close to the cathode, the farthest location from the PMTs. This study allowed a direct assessment of how the 20~PE threshold is expected to impact the acceptance in MicroBooNE. This is examined further in section~\ref{sec:trig_eff} using data taken at distances greater than 250 cm from the PMT planes. That sample includes events below the standard 20 PE threshold, which are typically excluded from MicroBooNE analyses.

\begin{figure}[h!]
    \centering
    \begin{subfigure}{0.8\textwidth}
        \centering
        \includegraphics[width=\linewidth]{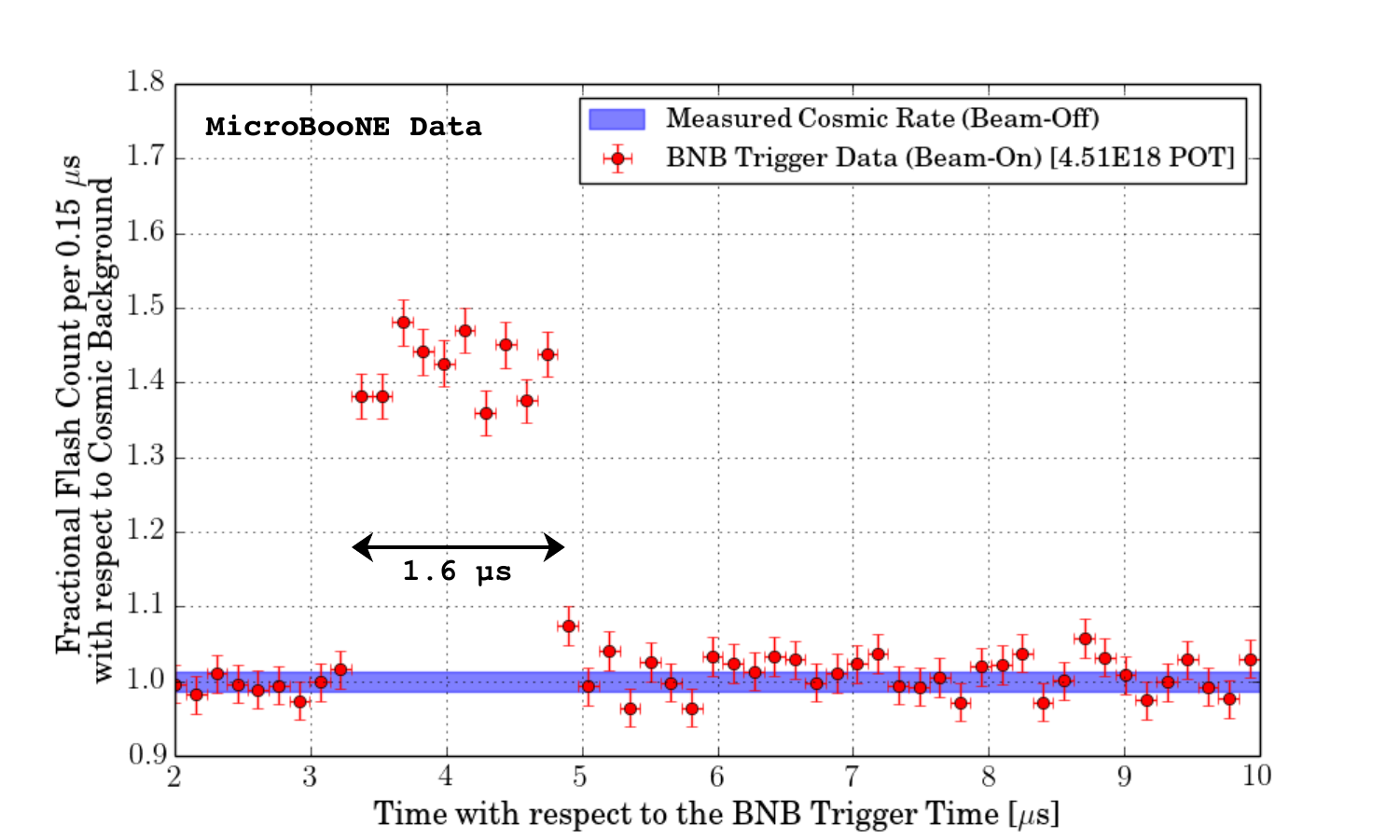}
        \caption{}
        \label{fig:bnb_numi_trigger_a}
    \end{subfigure}
    \vspace{1em}
    \begin{subfigure}{0.8\textwidth}
        \centering
        \includegraphics[width=\linewidth]{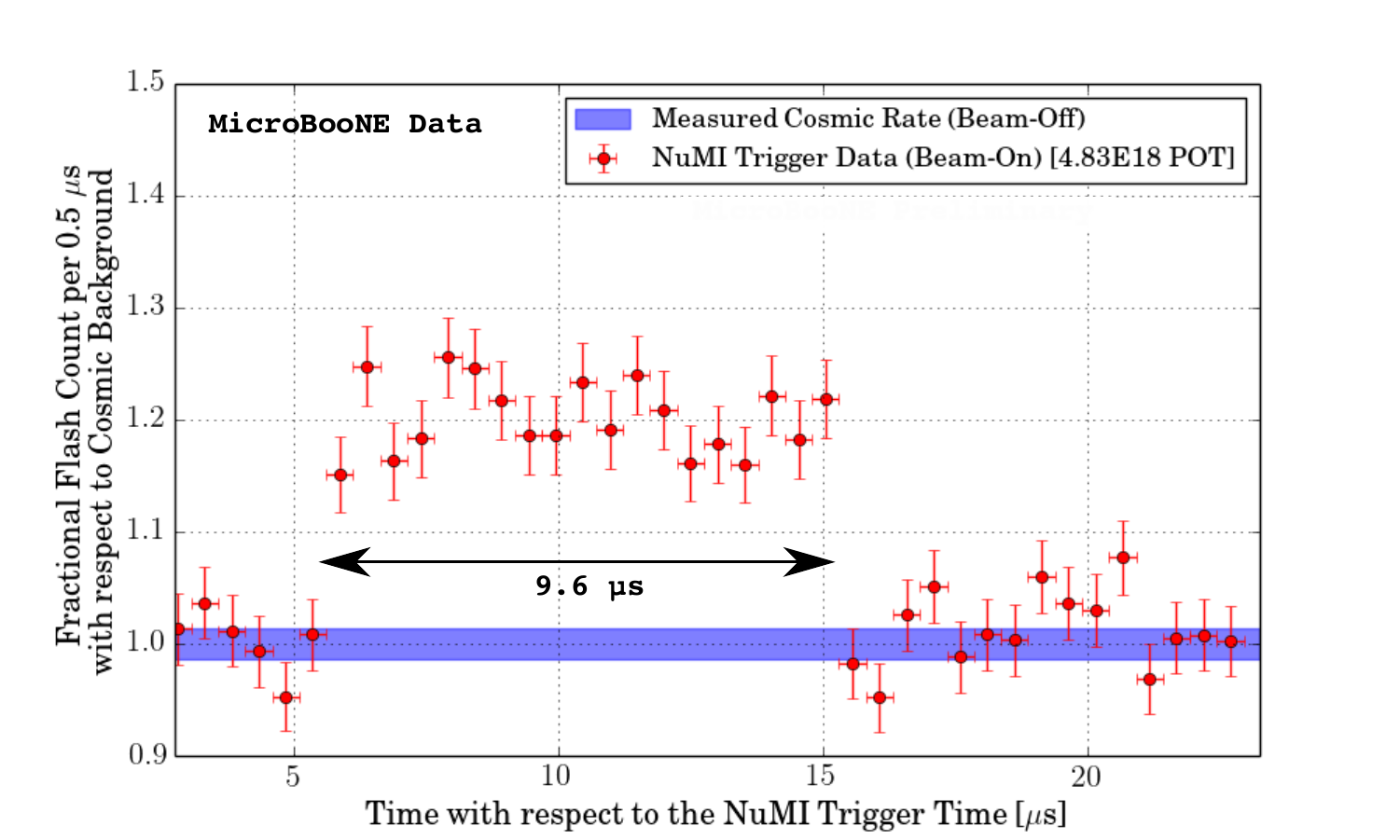}
        \caption{}
        \label{fig:bnb_numi_trigger_b}
    \end{subfigure}
    \caption{(a) Measured distribution of flash times relative to the trigger time for BNB-triggered events, shown as a ratio to the expected cosmic rate from off-beam data. (b) Same distribution for NuMI-triggered events. In both cases, the blue band indicates the expected cosmic rate (centred at 1) with its uncertainty. A clear excess from beam-induced events is visible in the range of 3-5~µs for BNB and 5-15~µs for NuMI.}
    \label{fig:bnb_numi_trigger}
\end{figure}

During the beam window, cosmic rays generate light at random times, but neutrino interactions produce a localised excess that is temporally aligned with the beam spills. Optical waveforms are processed to identify time-coincident light pulses (“optical hits”), which are grouped within 100~ns into “optical flashes,” each characterised by its time relative to the trigger and its total number of reconstructed PEs~\cite{MicroBooNE:2020vry, Caratelli:2020dlg}. The number of PEs per optical hit is estimated based on the integrated pulse area and the PMT gain. The total number of PEs in an optical flash is then the sum of all its optical hit PEs. Figure~\ref{fig:bnb_numi_trigger} shows the flash time distribution for BNB- and NuMI-triggered events, normalised to the expected cosmic rate measured from off-beam data (blue band centred at 1.0). To enhance the neutrino contribution here, only flashes with reconstructed PE between 50 and 1500 were included beyond the standard offline trigger threshold. A clear excess corresponding to the 1.6~µs BNB spill and 9.6~µs NuMI spill is observed. 

\begin{figure}[h!]
 \centering
 \includegraphics[width=1.0\textwidth]{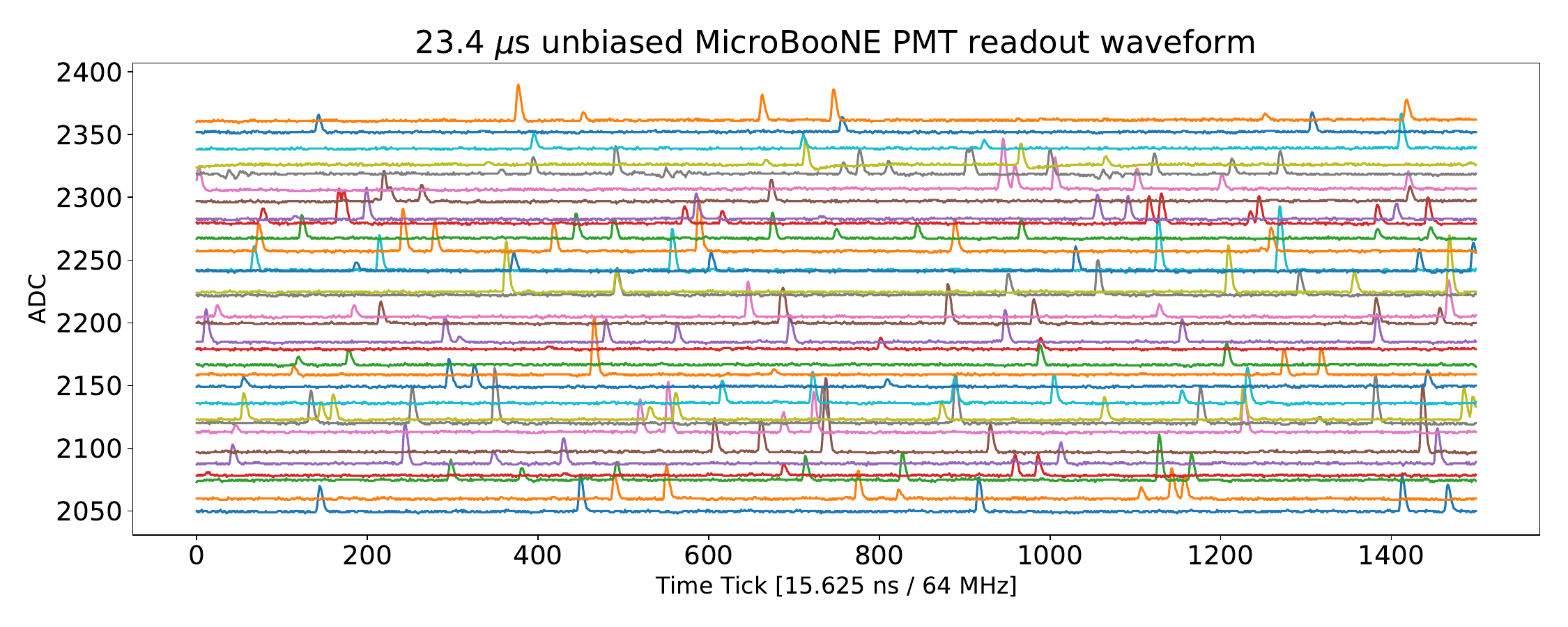}
 \caption{Example of 23.4~µs waveform in the MicroBooNE optical readout collected with a random trigger. The waveforms show individual SPE pulses on each PMT.}
  \label{fig:spe_waveforms}
 \end{figure}

MicroBooNE also applies random triggers to collect data outside the beam-gate window for use in various calibration studies. The light signals collected by random triggering are shown in figure~\ref{fig:spe_waveforms}. The small pulses seen randomly distributed in each waveform are single photoelectrons (SPEs). These are seen in figure~\ref{fig:beam_waveforms} outside of the large peak areas. MicroBooNE observed a random SPE rate $\mathcal{O}$(200~kHz) per PMT as elaborated on in section~\ref{sec:spe_rate}, which is consistent with preliminary rates found in LArIAT at Fermilab~\cite{Priv_Comm_lariat}, and ProtoDUNE at CERN~\cite{lidine_dante_spe}, accounting for different active argon volumes and light detectors. The continuous high rate of SPEs provides a valuable PMT calibration source, as discussed in section~\ref{sec:pmt_gain}. This avoids the need for dedicated data collection for calibrations that would interrupt beam data collection. 

\section{PMT gain measurement and calibration}{\label{sec:pmt_gain}}
PMT gain is measured based on the recorded charge and amplitude of optical pulses resulting from single photoelectrons (SPEs) being observed by a PMT. PMT gains are prone to change and need to be re-calibrated every time the HV system is powered off for a prolonged time (generally a few times per year for detector component maintenance or during building power outages)~\cite{Saia:2024adm}. Additionally, cryogenic PMT gains have been observed to occasionally drift with time, which also requires re-calibration. This is done by re-calculating the PMT gain and, if necessary, applying minor adjustments to the high voltage bias with a target of keeping the gains around 20~ADC/PE $\pm$10\% for the majority of the PMTs. This requirement was put in place after the first year of running, after significant deviations from this baseline gain were observed in multiple PMTs.

The observed charge, defined as the baseline-subtracted integral under a segment of a PMT waveform, is proportional to the incident photon flux, the quantum efficiency of the photocathode, and the photomultiplier gain. Gain is often expressed in charge recorded per incident photoelectron, or ADC/PE.  

It is common to employ the use of a laser or an LED calibration light source to generate variable light intensity to calibrate the time-response as well as gain of PMTs. MicroBooNE installed and employed a flasher LED system~\cite{Conrad:2015xta} that was used during early detector commissioning. However, for the physics data, the random, diffuse background SPEs present in PMT signals are used to determine gain calibrations. The $\mathcal{O}$(200~kHz) SPE rate observed in MicroBooNE leads to $O$(5) unique pulses per 23.4\,$\upmu$s of unbiased waveform per PMT. An example of this can be seen in figure~\ref{fig:spe_waveforms} where pulses within the waveforms are SPE candidates. The use of these random SPEs obtained in the readout window has significant advantages: the pulses are uniformly distributed in time throughout the detector volume; there is clear separation between the pulses; and it does not require a dedicated readout configuration in the DAQ, eliminating the need to interrupt valuable physics data-taking. This resulted in an accumulation of a continuous dataset for the full data-taking period to perform the PMT gain calibration. This dataset comes from an already-running data stream, described earlier, in which TPC and PMT event readouts are randomly triggered outside the beam window with no requirement on the minimum amount of light observed by the PMTs. This enabled an unbiased triggering scheme which was an important requirement for gathering low-light pulses to properly determine the PMT gain.

\subsection{Pulse finding and selection}

SPE pulses are selected based on pulse amplitude, integrated charge, baseline stability, and isolation from larger beam- or cosmic-induced pulses. The extraction of SPEs from the digitized, baseline-subtracted high gain waveforms is performed using a constant fraction discriminator (CFD) algorithm. The CFD identifies pulse start and end times and amplitudes by triggering at a fixed fraction of each pulse’s maximum height rather than at a fixed absolute threshold. This approach ensures consistent timing and amplitude selection for pulses of identical shape but varying size, as expected for single photoelectrons. In this implementation, the CFD threshold corresponds to 10~ADC, which is approximately 0.5~PE, or about twenty times the typical root-mean-square (RMS) of the baseline noise. This effectively separates the pedestal (0~PE) from the first PE peak without biasing the pulse selection toward higher amplitudes.

To isolate SPE pulses, several selection criteria are applied to identified pulses. The pulse amplitude and area are required to remain below 50~ADC and 500~ADC~×~time-ticks, respectively, ensuring that the selected pulses fall within the one to two PE region. In addition, the baseline RMS before and after each pulse must be less than 2~ADC and the ratio of the baseline levels before and after the pulse is required to lie between 0.95 and 1.05. These conditions suppress baseline fluctuations and reject overlapping or distorted pulses, ensuring the selected pulses are unaffected by ringing or ripples in the baseline of the PMT induced by bright flashes of light from cosmic muon events. These 1~PE pulses are then used to extract the gain for each PMT.

\subsection{Multi-photoelectron fitting for gain extraction}

The observed amplitude and area distributions of the selected SPE candidate pulses are fitted to extract the gain values for each PMT. The selected pulses correspond primarily to one- and two-PE observed. 
The number of PE to reach the first dynode of a PMT, $n$, for an average number of detected PE $\lambda$ is described by a Poisson distribution:
\begin{equation}
    P(n;\lambda) = \frac{\lambda^n e^{-\lambda}}{n!}\label{eq:1}.
\end{equation}
Given that the amplification of the first dynode is large, the dynode chain amplification ($G(x)$) is Gaussian:
\begin{equation}
    G(x) = \frac{1}{\sqrt{2\pi\sigma^2}}e^{-\frac{(x-\mu_{\mathrm{SPE}})^2}{2\sigma^2}}\label{eq:2},
\end{equation}
where $x$ is the measured pulse area (representing recorded charge), $\mu_{\mathrm{SPE}}$ is the mean of a 1-PE pulse area, and $\sigma$ is the standard deviation of a 1-PE pulse.
Combining equations \ref{eq:1} and \ref{eq:2} , the observed distribution of pulse area for $n$ PE, $S(x)$, becomes:
\begin{equation}
    S(x) = A\sum_{n=1} \frac{\lambda^n e^{-\lambda}}{n!}\frac{1}{\sqrt{2\pi (n\sigma)^2}}e^{-\frac{(x-n\mu_{\mathrm{SPE}})^2}{2(n\sigma)^2}},
\end{equation}
where $A$ is a normalization factor accounting for varying statistics in each fit.

For one-PE pulses or multi-PE pulses where the PEs arrive at the same time, the total SPE pulse amplitude will follow the same relationship, and exhibit a linear correlation with total pulse area. However, it should be noted that if the observed SPE candidates comprise several PEs offset in time within the SPE window, the amplitude will not scale the same way as the area. As the majority of selected pulses are in fact 1-PE, in practice MicroBooNE uses the same analytical form to describe both the SPE pulse area and amplitude distributions, with $\mu_{\mathrm{SPE}}$ being the mean, and $\sigma$ being the standard deviation of either the 1-PE pulse area or amplitude, respectively.   
The extracted area gain is used in calibration to scale data and MC reconstructed optical PE, while the amplitude gain is used as a proxy for the stability of PMT gains during detector operations. 

\begin{figure}[h!]
    \centering
    \begin{subfigure}{0.6\textwidth}
        \centering
        \includegraphics[width=\linewidth]{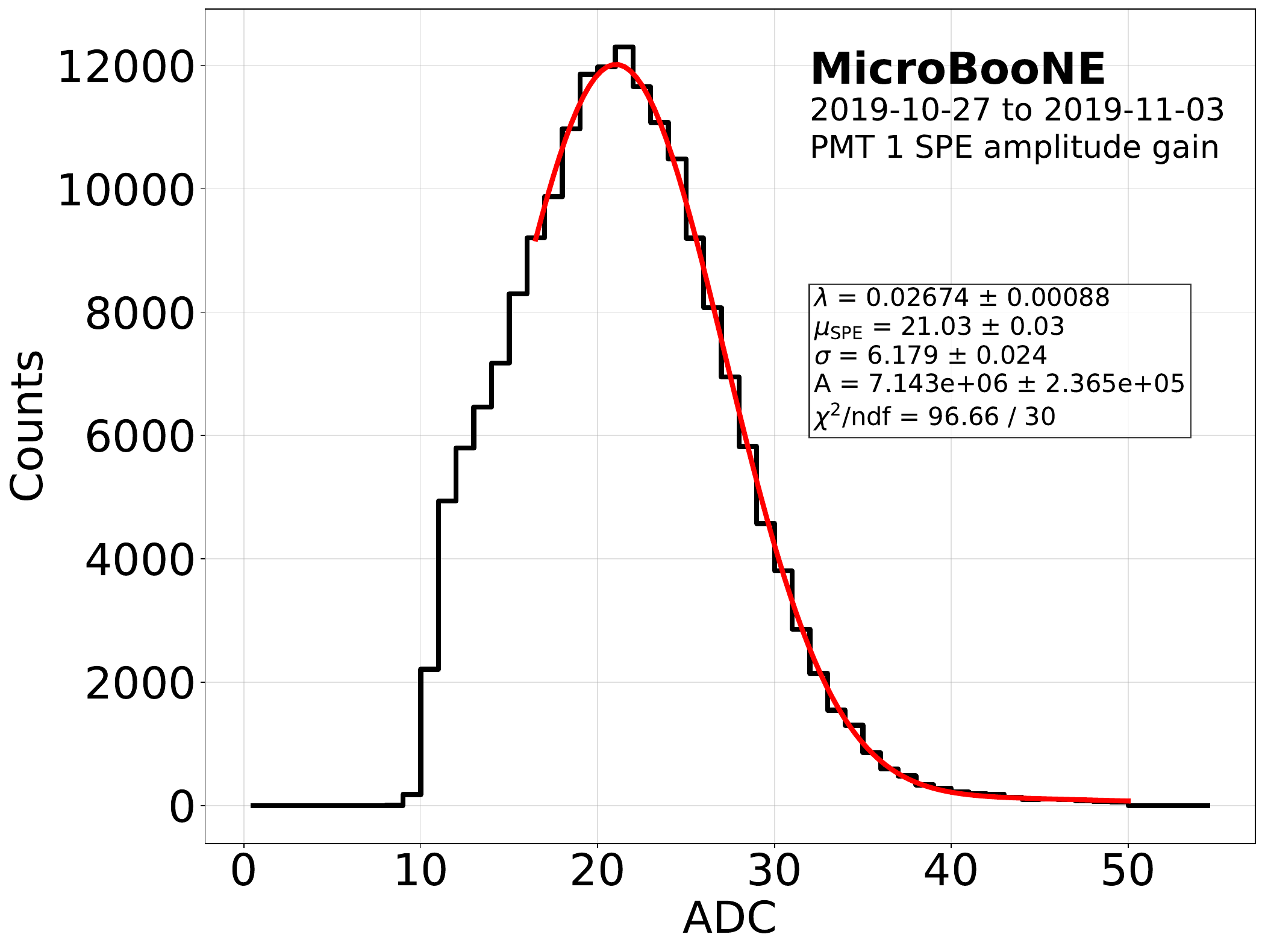}
        \caption{}
        \label{fig:spe_fit_a}
    \end{subfigure}
    \vspace{1em}
    \begin{subfigure}{0.6\textwidth}
        \centering
        \includegraphics[width=\linewidth]{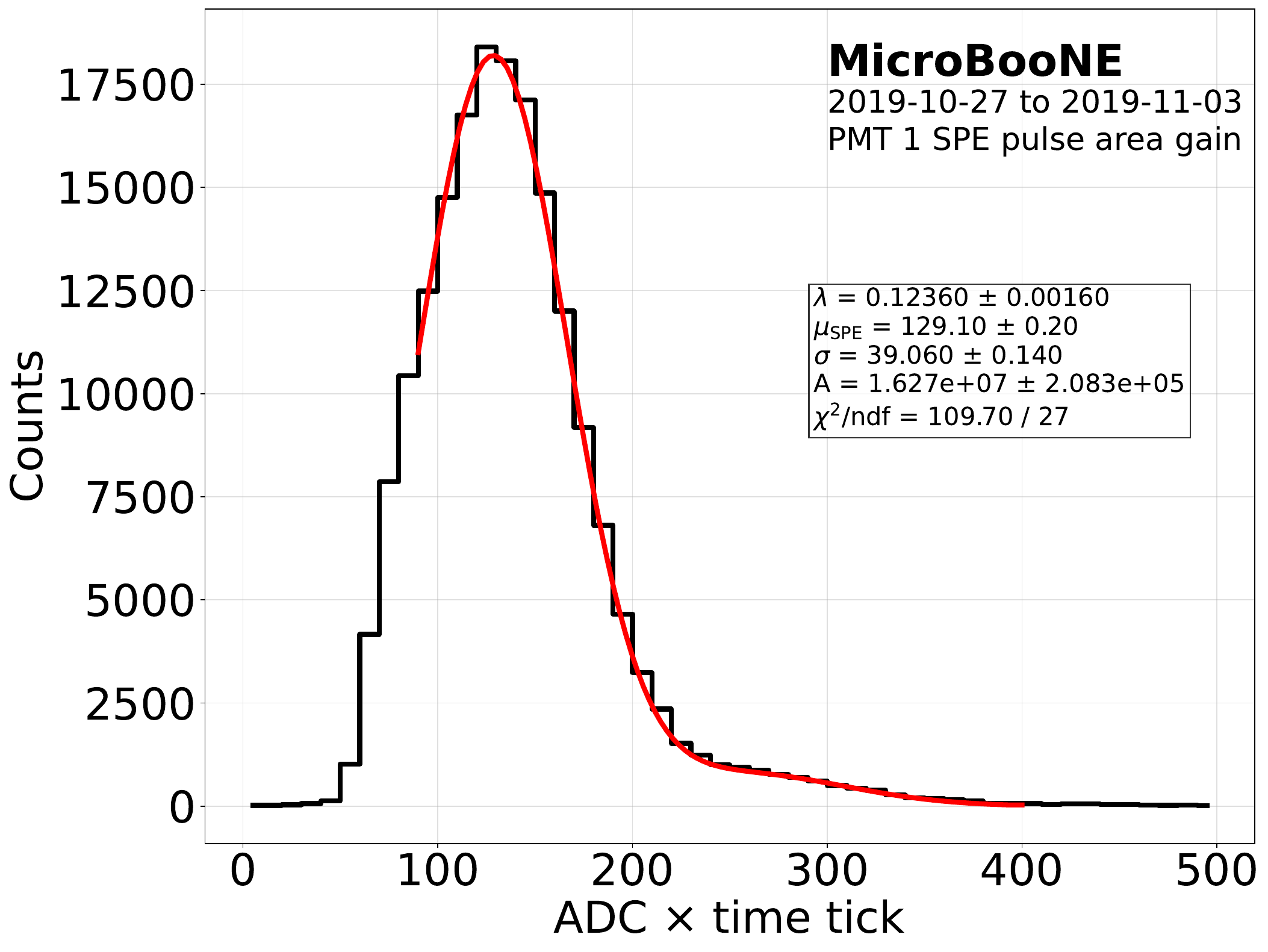}
        \caption{}
        \label{fig:spe_fit_b}
    \end{subfigure}
    \caption{(a) Example SPE pulse–amplitude fit and (b) SPE pulse–area fit for one PMT channel. The fitted parameter $\mu_{\mathrm{SPE}}$ represents the effective gain used in the calibration.}
    \label{fig:spe_fit}
\end{figure}

The above analytical form is fitted to the cumulative distribution of pulse amplitude and pulse area for all selected pulses to extract the average amplitude (area) parameter $\mu_{\mathrm{SPE}}$ which is an effective gain calibration factor. In practice, given the above assumption that one- and two-PE pulses pass the selection criteria, the number of observed PE $n$ is limited to 2 in the fit.
An example of analytical fitting to the pulse amplitude and area distributions for one PMT channel is shown in figure~\ref{fig:spe_fit}.

\subsection{PMT gain calibration}

\begin{figure}[h!]
 \centering
 \includegraphics[width=0.9\textwidth]{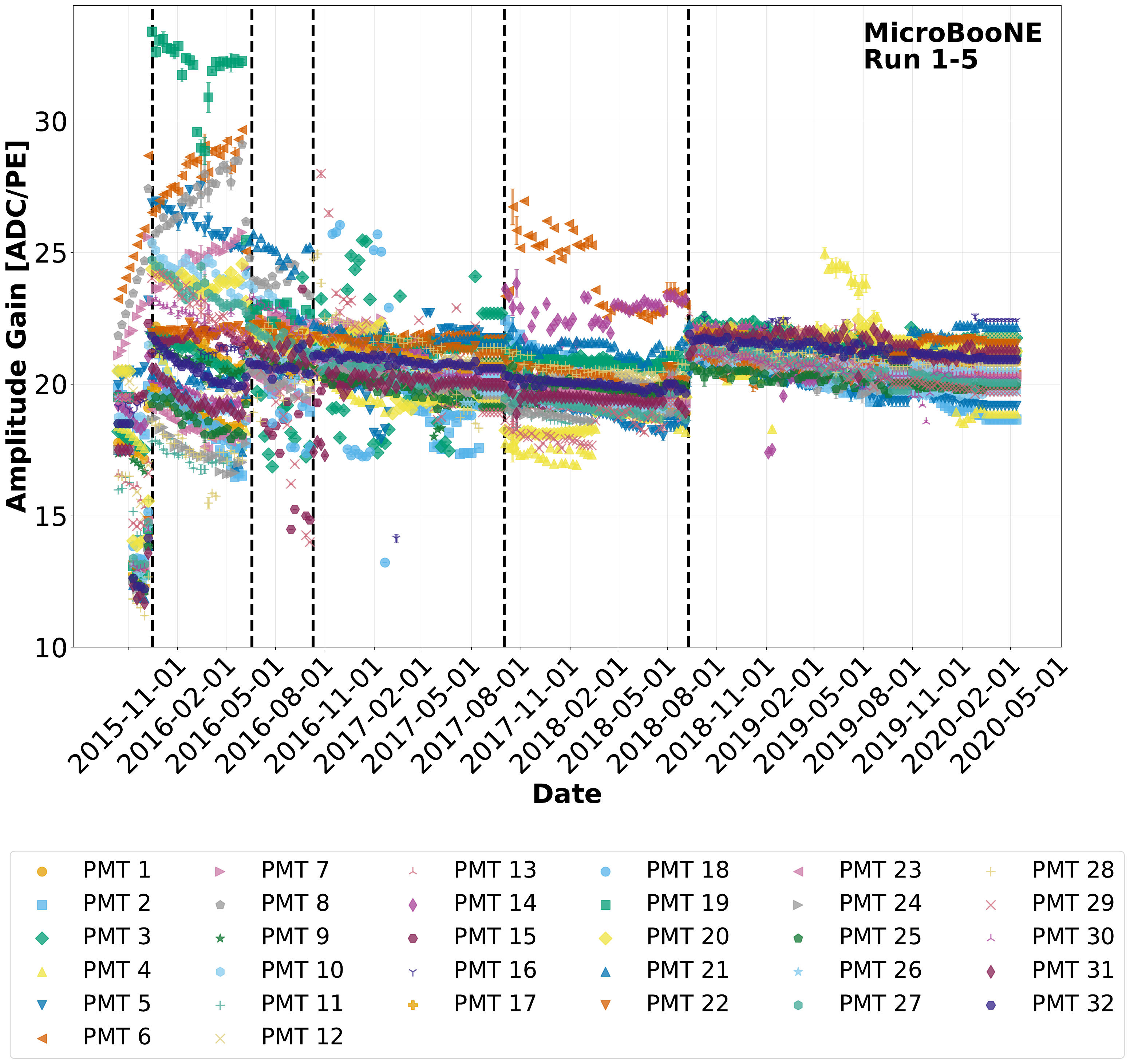}
 \caption{Evolution of each of the PMT gains from SPE pulse amplitude over the full data taking period. The vertical dashed lines represent major changes to the PMT HVs throughout operation.}
  \label{fig:gain_evo}
 \end{figure}

Operationally, the PMT gain is measured with the above procedure for the full MicroBooNE dataset divided into one-week periods. The evolution of measured PMT gains from pulse amplitude is shown in figure~\ref{fig:gain_evo} for each PMT using the fitted values of $\mu_{\mathrm{SPE}}$. During early data collection there was significant spread in the PMT gains and sizeable drifting over time. This spread was one of the reasons for setting the offline trigger threshold at 20~PE, well above the level where gain variations affect the trigger decision. The sharp steps (mid September 2015, mid March 2016, mid July 2016, start of July 2017, start of June 2018) are a result of adjustments to the PMT HV that were made to adjust the gains to nominal. The vertical dashed lines represents times when significant HV changes occurred on a large subset of the PMTs, with the last three corresponding to the end of the regular beam operation during summer shutdowns. Other minor adjustments to single PMTs were done as necessary (e.g. PMT 4 in May 2019). After an HV bias power supply was replaced during the summer of 2017, most PMT gains were stable. There was a final major adjustment to the HV in summer 2018 for the remainder of operations. Periods of constant gain values for all PMTs (e.g. June-July 2017, July 2019, January-March 2020) are associated to periods when the PMTs were powered off due to operational tasks or refurbishment, or when the argon purity was extremely poor.

The extracted gains are recorded in a calibration database with their associated time periods of validity (typically one week unless insufficient statistics required a longer period), providing a record of both routine fluctuations and significant adjustments to the PMT high voltage. This procedure ensures that all analyses in MicroBooNE incorporate the relevant gain information for the full time period. The measured gains serve a dual purpose: they correct the optical reconstruction in data to ensure uniformity across all PMTs, and they are applied to Monte Carlo (MC) simulations to bring the simulated optical response into closer agreement with the observed data. By default, optical simulation is run with mean SPE area of 120~ADC~$\times$~time tick and mean amplitude of 20\,ADC while the measured PMT gains vary, as shown in figure~\ref{fig:gain_evo}. Fluctuations in PMT gain are normal and arise from several factors, including intrinsic differences between individual PMTs, variations in operating voltage, and changes in response linearity over time. Such behaviour is typical for PMT systems and underscores the importance of regular calibration to maintain consistent detector response. The measured pulse area gains are used as a multiplicative factor of $120/\mu_{\mathrm{SPE}}$ to scale the number of reconstructed PE in each optical flash in data and MC to get closer agreement. 

\begin{figure}
    \centering
    \includegraphics[width=0.5\linewidth]{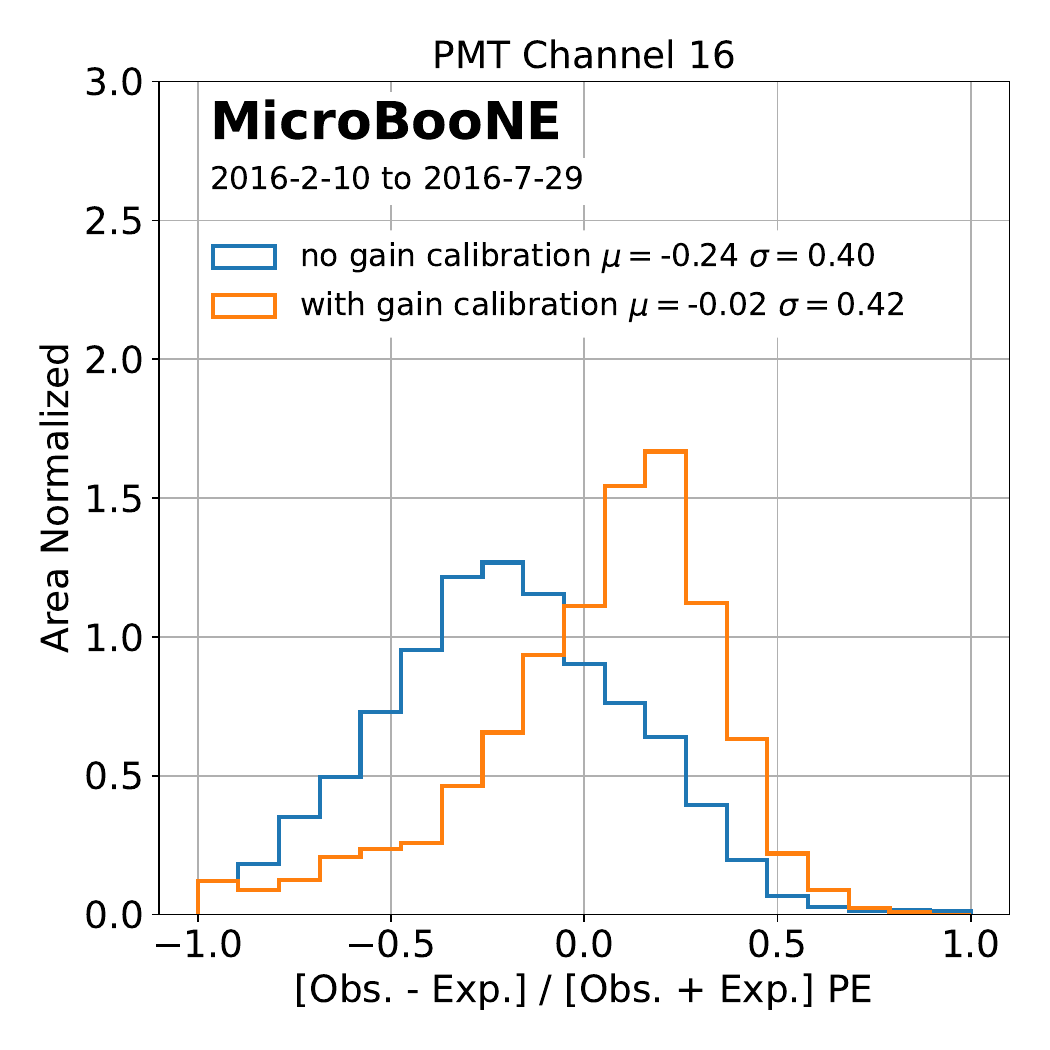}
    \caption{Impact of the gain pulse area calibration for a typical PMT using a sample of cathode piercing tracks collected over Run 1. The gain calibration shifts the distribution mean $\mu$ reducing the overall residual difference between the PE hypothesis (Exp.) and reconstructed optical flash PE (Obs).}
    \label{fig:gain_corr_phys}
\end{figure}

The effect of the gain pulse area calibration on physics events, which are typically associated with multi-PE optical flashes, is shown in figure~\ref{fig:gain_corr_phys} for a typical PMT. For a sample of cathode-piercing muon tracks (see section~\ref{LY_sec} for the selection criteria and description) from MicroBooNE Run 1 (Feb.\ 10, 2016 - Jul.\ 29, 2016), the number of reconstructed optical-flash PE with and without the gain correction applied is compared to the predicted number of PE given the TPC track energy and path (called the PE hypothesis). An additional requirement that the total reconstructed optical‐flash PE seen by the PMT is between 30 and 200 approximately represents the typical amount of light seen from neutrino interactions and removes poorly reconstructed pulses with very low or very high charge. The correction reduces the difference between data and hypothesis.

\section{Time-based light yield stability}{\label{LY_sec}}

It is important to track the performance of the light detection system and account for any time-dependent effects such as changing amounts of impurities in the argon, ageing of the PMTs, and degradation of the TPB wavelength shifter. While it can be very difficult to disentangle these various effects, a targeted calibration campaign can minimise the impact of these factors on analysis results. A metric for the performance of the light detection system is the light yield. This is defined as the total number of PE detected across all of the PMTs per unit of energy deposited in the detector. The stability of the light yield then can be assessed over time and any change used as a time-dependent correction in simulation.

\subsection{Data sample and selection}

Since MicroBooNE is located on the earth's surface, it is exposed to a high flux of cosmic-ray muons. This enables a high-statistics time-dependent measurement of the light response. The measurement uses samples of through-going cosmic-ray muon tracks that cross either the anode or the cathode of the detector, referred to as anode or cathode piercing tracks (ACPT). These samples are useful for the light calibration since they are minimally dependent on the light for reconstruction. Tracks crossing the anode or the cathode of the detector can be identified through the charge information alone based on the expected position of the anode or cathode in the drift direction relative to the trigger time~\cite{MicroBooNE:2020kca}. This minimises any biases that could occur during reconstruction if the light response varies over time.

\begin{figure}
\centering
\includegraphics[width=0.5\textwidth]{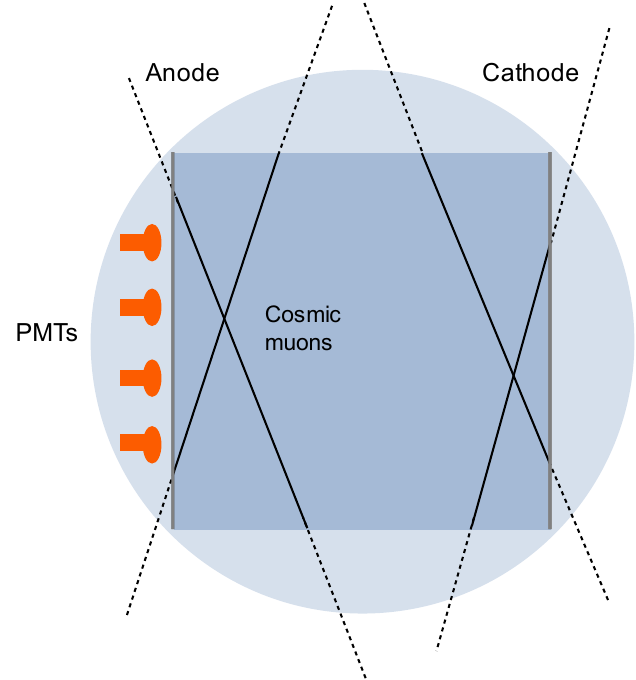}
\caption{Illustration of the cosmic-ray-induced muon tracks selected in the anode piercing and cathode piercing samples.}
\label{fig:acpt_diagram}
\end{figure}

\begin{figure}
\centering
\includegraphics[width=0.7\textwidth]{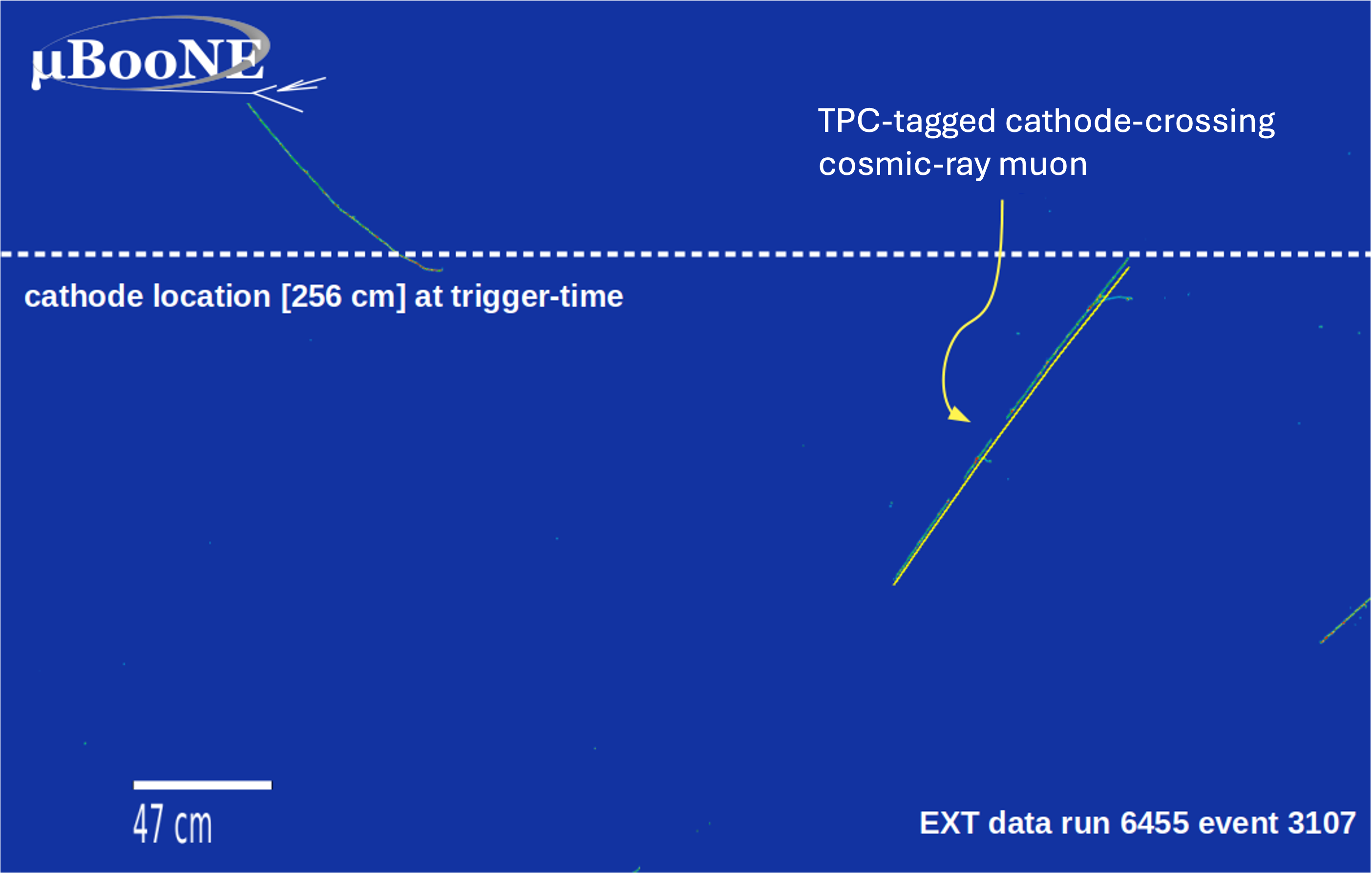}
\caption{Event display of an example selected cathode piercing track. The deposited charge (colour scale) is shown as a function of readout wire number (x-axis) and time (y-axis). The white dashed line shows the position of the cathode. The yellow line shows the reconstructed cathode piercing track.}
\label{fig:acpt_event_display}
\end{figure}

The ACPT samples are divided into two sets, each illustrated in figure~\ref{fig:acpt_diagram}, according to whether they are anode-piercing (APT) or cathode-piecing (CPT). Tracks are selected that either enter the top of the detector and exit through the anode/cathode, or enter the anode/cathode and exit through the bottom of the detector. An event display of an example cathode-piercing track is shown in figure~\ref{fig:acpt_event_display}, where the position of the cathode is shown by the dashed white line. The selected tracks are distributed throughout the entire length of the detector in the beam direction and have no threshold applied to the individual PMT readout. 

Once selected using the TPC, optical flash matching is performed to identify the associated light signal and improve the purity of the sample. The average position of the flash along the beam direction is evaluated based on the positions of the PMTs that detect light, weighted by the number of photons. This is required to be within 100 cm of the centre of the track measured along the beam direction. Next, several quality requirements are applied in order to remove poorly reconstructed tracks and outliers. Tracks are removed that have: length $L < 40$\,cm or $L > 400$\,cm; total number of photons $N_{PE} > 10000$ for APT or $N_{PE} > 1000$ for CPT; or that start or end within 50\,cm of the cathode for APT or anode for CPT.

The ACPT sample spans the time period October 2015 to March 2020, covering the full physics data taking period along with the summer beam shutdowns between each run. The event rate varies slightly over time due to two factors: larger samples are collected during the summer beam shutdowns; and a larger fraction of recorded events are reconstructed during later run periods due to increased availability of computing resources. Once data quality requirements are applied, we select approximately 900,000 anode-piercing tracks and 600,000 cathode-piercing tracks in total. The sample of selected APT is larger than the CPT since the APT pass closer to the PMTs, resulting in more light being collected and correspondingly a higher probability of triggering the detector. 

Two periods of low argon purity are also removed from the samples. These span from September 22nd to October 6th 2018 and from February 20th to March 1st 2020. During these periods the electron lifetime was insufficient for the ionisation charge to drift from the cathode to the anode. This results in the APT appearing shorter, since only the charge close to the anode is seen, resulting in them appearing to have a disproportionately large light yield relative to the visible ionisation charge. In addition, during these periods no CPT are reconstructed. 

\subsection{Method and results}
\label{sec:lydecline}

The selected through-going muon tracks are approximately minimally ionising for the entirety of their path in the detector. The effective light yield of each muon can therefore be evaluated in terms of the total number of photons observed across all PMTs divided by the length of the muon track, in PE/cm. This serves as a proxy for the number of photons per unit of energy deposited, without needing to depend on the calibration of the ionisation charge collection system. 

Since the PMTs are located only at the anode of the detector, the light yield is highly non-uniform in the drift direction. This is further compounded by the effects of Rayleigh scattering that impedes the propagation of light from the cathode to the PMTs~\cite{Babicz:2020den}. Therefore, significantly more light is observed for anode-piercing tracks than cathode-piercing tracks. To allow comparison between the two sub-samples, the relative change in the light yield is assessed over time rather than the absolute yield.

\begin{figure}[h!]
    \centering
    \begin{subfigure}{0.495\textwidth}
        \centering
        \includegraphics[width=\linewidth]{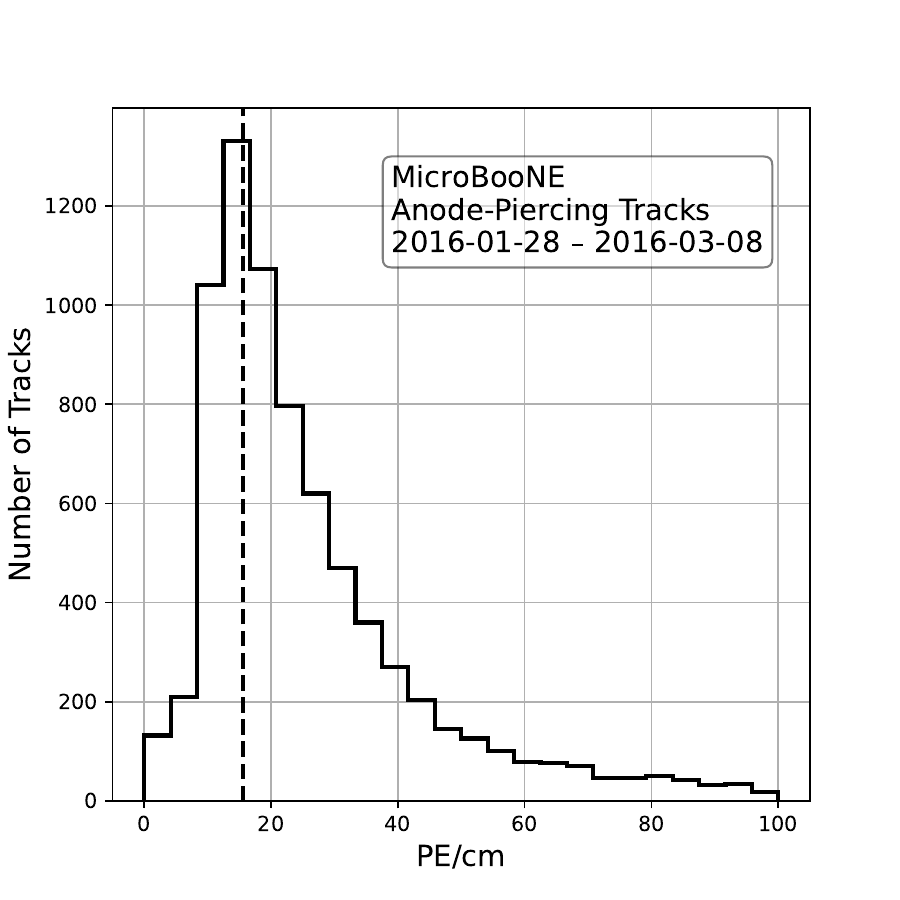}
        \caption{}
        \label{fig:truncated_median_a}
    \end{subfigure}
    \hfill
    \begin{subfigure}{0.495\textwidth}
        \centering
        \includegraphics[width=\linewidth]{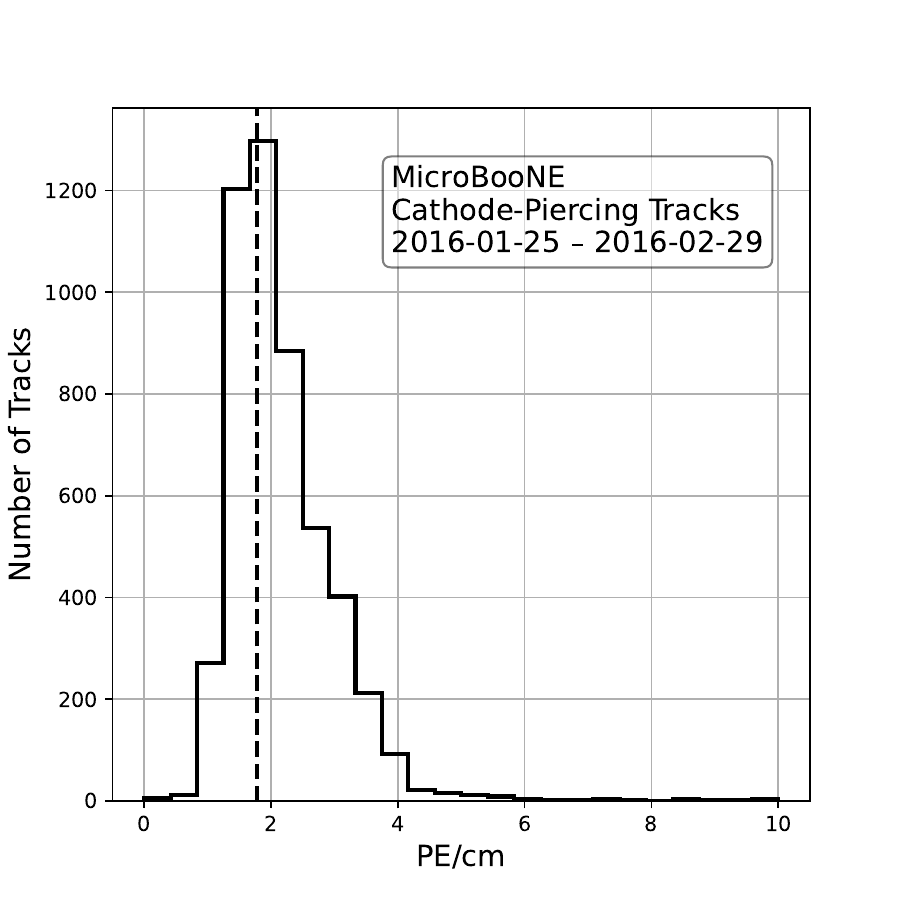}
        \caption{}
        \label{fig:truncated_median_b}
    \end{subfigure}
    \caption{Distribution in PE/cm for (a) anode-piercing tracks and (b) cathode-piercing tracks, shown for a single example time bin. The light yield is quantified using the truncated medians of the distributions, indicated by the vertical dashed lines.}
    \label{fig:truncated_median}
\end{figure}

The APT and CPT samples are divided into bins in time. To reliably assess the variation of the light yield over time, the bin width is allowed to vary to ensure the same number of tracks is present in each bin. Anode piercing tracks pass through the side of the detector where the PMTs are located. This can result in larger variation in the light yield between tracks, with tracks passing very close to a particular PMT having a higher yield than tracks passing between two PMTs. To reduce the impact of this, a larger sample is used per bin for the APT (7500 $\text{tracks}/ \text{bin}$) compared with the CPT (5000 $\text{tracks}/ \text{bin}$). This also accounts for the differing total sample sizes giving a similar number of bins overall for each classification. The light yield in each bin, in PE/cm, was then quantified by evaluating an iterative truncated median of the distribution trimming values more than 2$\upsigma$ from the median until convergence. This allows the peak of the distribution to be identified without being skewed by outlier events. Examples of the distribution within a single time bin and the iterative truncated median algorithm being applied are shown in figure~\ref{fig:truncated_median} for APT (a) and CPT (b).

\begin{figure}
\centering
\includegraphics[width=0.95\textwidth]{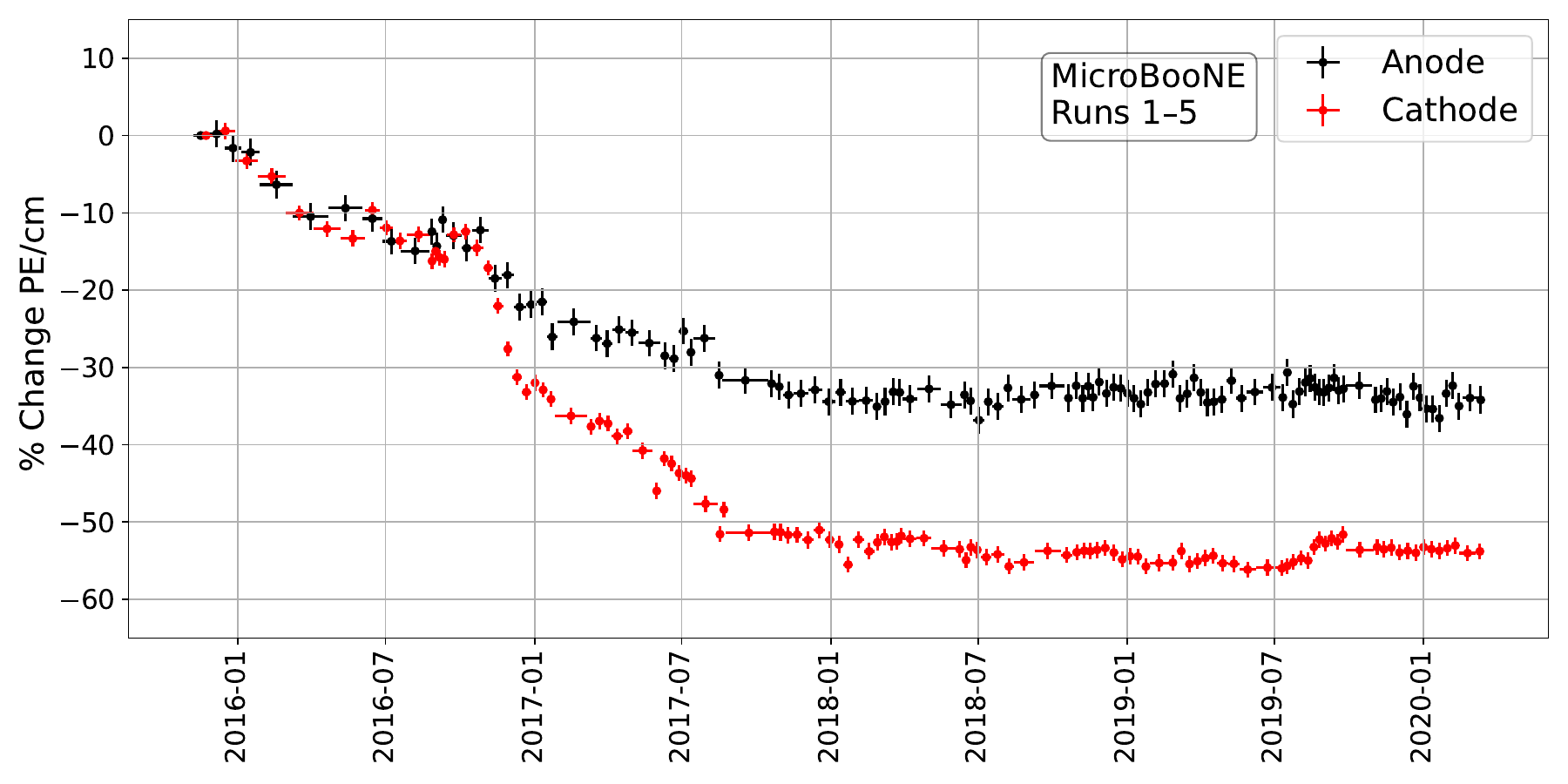}
\caption{Percentage change in the light yield over time for the anode piercing tracks (black) and cathode piercing tracks (red). The vertical error bars represent the statistical uncertainty and the horizontal error bars show the bin width.}
\label{fig:light_response_change}
\end{figure}

The percentage change in the light yield over time is shown in figure~\ref{fig:light_response_change} for the anode-piercing tracks (black) and cathode-piercing tracks (red). The light yield is observed to decline significantly between the start of data-taking and approximately January 2018 for both the anode-piercing and cathode-piercing tracks. It then remains approximately stable until the end of MicroBooNE's physics runs in March 2020. The amplitude of the decline is significantly different for the anode-piercing and cathode-piercing tracks: approximately 35\% at the anode, and 55\% at the cathode. The decline initially occurs at a similar rate for both types of tracks until late 2016, at which point the decline accelerates and the cathode-piercing tracks see a significantly larger decline than the anode-piercing tracks. Finally in August 2019 a slight increase in the light response is seen, particularly for the cathode-piercing tracks. This corresponds with a filter regeneration that occurred during August 2019. 

The cause of the light response decline is unknown. Possible explanations include degradation of the PMTs, degradation or loss of the wavelength shifter coating the plates mounted in front of the PMTs, or the introduction of a contaminant into the detector that is not removed by the filtration system. These are currently the subject of active investigation and are discussed in more detail in section~\ref{sec:ly_decline_discussion}. 

The difference in the size of the light response decline between the anode and cathode piercing tracks is also unclear in origin. This effect is difficult to study further in this analysis due to the extended tracks that the muons produce, and the fact that the selected muons deposit significant energy outside of the active volume of the detector, external to the TPC. Future analysis using alternative samples with more localised energy depositions, such as isolated protons or Michel electrons, may provide further insight into these effects.  

To assess the impact of Cherenkov radiation on the difference in the light response decline between anode and cathode piercing tracks, the APT and CPT samples were each divided into two subsamples: tracks entering the top of the detector and exiting the anode/cathode, and tracks entering the anode/cathode and exiting the bottom of the detector. No statistically significant difference was observed in the light yield decline between the two subsamples, suggesting that Cherenkov radiation has a negligible impact on this measurement. 

\subsection{Light yield calibration}
\label{sec:calibration}

Since the light response varies substantially over time, a time-dependent calibration is required to model it. Since the cause of the light response decline and the difference between the anode and the cathode are unknown, the calibration values are taken as the midpoint between the APT and CPT measurements with an uncertainty corresponding to half of the difference between them. This avoids the need to assume a particular model for the anode/cathode difference, such as absorption, and instead allows the range of the decline to be evaluated as a systematic uncertainty in analyses. 

The resulting calibrations are stored in a calibration database with approximately two week intervals. They are applied to correct the number of photons in MC simulation to model the change in the light yield over time and allow any impact to be taken into account.

\section{Scintillation light triggering efficiency measurement}{\label{sec:trig_eff}}

To assess the impact of the MicroBooNE trigger on low-energy neutrino interactions, it is essential to measure the scintillation light triggering efficiency in this energy range. Figure~\ref{fig:cv_library} shows the light yield as a function of drift distance and height in MicroBooNE, illustrating non-uniformities along the drift direction caused by the combined effects of light collection geometry and propagation in liquid argon. Events near the cathode ($x \simeq 250$~cm) have the lowest detected light, leading to reduced triggering efficiency and setting the lower bound on the detector’s triggering response.

The analysis in this section presents the scintillation light triggering efficiency of the MicroBooNE light detection system for prompt light (first 100~ns) as a function of the reconstructed visible energy deposited in the TPC using cathode-piercing cosmic-ray muon tracks tagged by the CRT. The focus is on the $\mathcal{O}(100~\mathrm{MeV})$ range, at the 20~PE threshold relevant for MicroBooNE’s software trigger. This threshold is applied offline to events recorded with the $\sim$5~PE online trigger. Events below 20~PE, but above 5~PE, remain in the recorded dataset.

\subsection{Samples and selection}

The event sample is defined using CRT tracks in coincidence with TPC ionisation charge, without imposing any light-based selection. A temporal match between CRT tracks and the unbiased PMT readout window ensured full light coverage for evaluating the impact of a 20~PE trigger threshold. Data from MicroBooNE Run~3 (early 2018) is used, corresponding to a period of stable light yield and active CRT operation, enabling CRT-TPC matching based solely on external timing. A dedicated dataset retains events below the nominal 20~PE threshold for an unbiased efficiency measurement.

High-statistics, topologically simple cosmic-ray muon events located near the cathode, which is the region farthest from the PMTs, are selected to probe the lowest light regime. These non-stopping cathode-piercing tracks were matched to in-time CRT hits to provide calorimetric information independent of reconstructed light signals. A corresponding simulated sample of 1–10~GeV cosmic-ray muons is generated and processed through the full MicroBooNE simulation and reconstruction chain. Both data and simulation underwent identical selection criteria, including CRT–TPC position and time matching, geometrical constraints near the cathode, and the removal of short or poorly reconstructed tracks.

\begin{figure}
	\centering
	\includegraphics[width=0.65\textwidth]{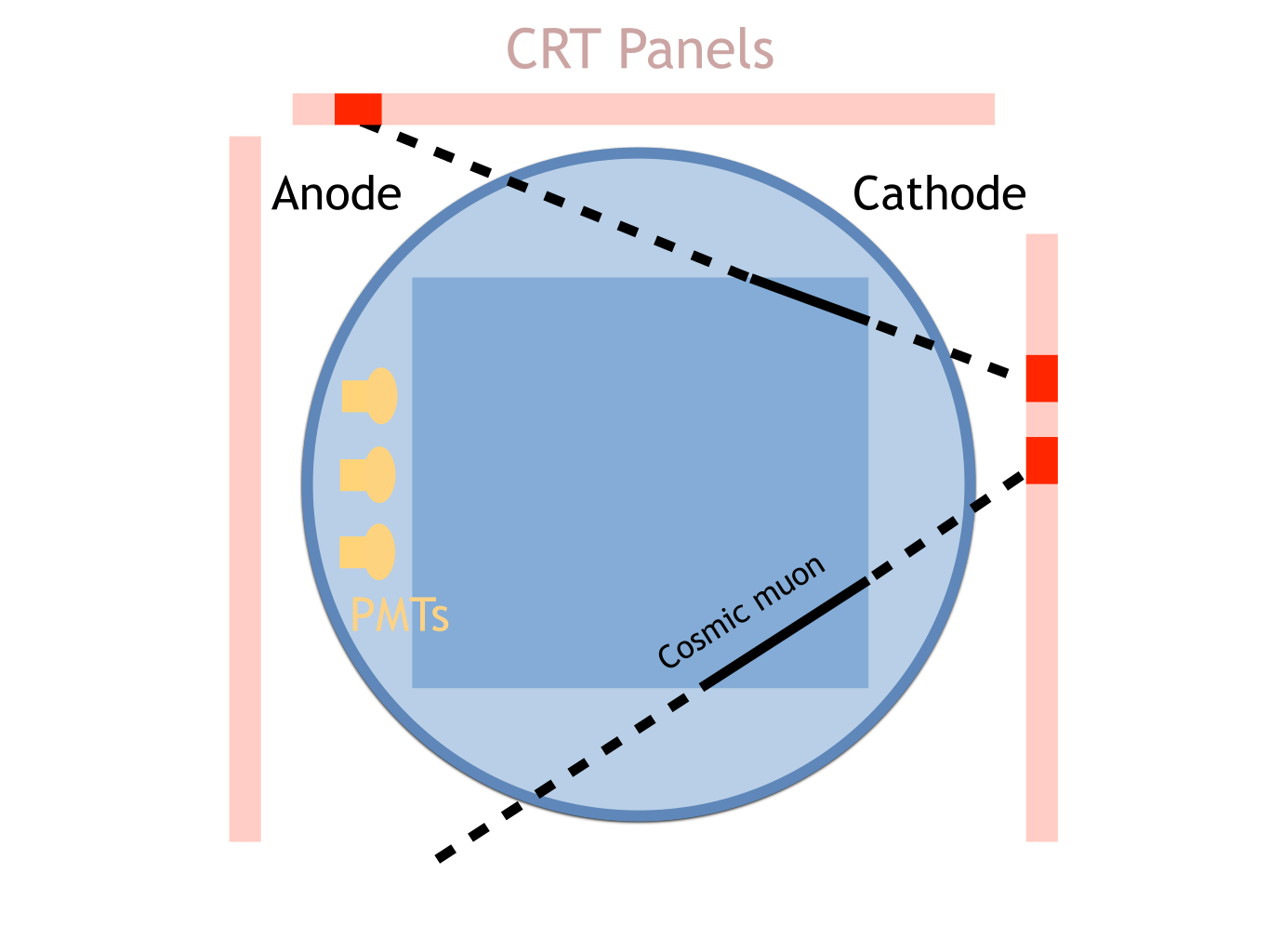}
	\caption{Cartoon of an example cosmic muon with trajectories that cross the cathode and are selected by the CRT.}\label{fig:sample_cartoon}
\end{figure}

\paragraph{CRT-TPC matching:}
Cathode piercing tracks are identified using TPC information as described in section~\ref{LY_sec}. Then the initial interaction time, $t_0$, is determined by the timing of coincident hits in the CRT~\cite{MicroBooNE:2021icu}, rather than using a flash matching technique. This avoids the need to use the $t_0$ from the PMTs, thus making the selection independent of light signals. After being $t_0$-tagged, the reconstructed TPC track is projected as a straight line towards the top and side CRT planes by taking advantage of the start and end positions of the reconstructed tracks as illustrated in figure~\ref{fig:sample_cartoon}. If the extrapolated line intersects the CRT hit in ($x$, $y$, $z$) within a 15~cm maximum distance at the same $t_0$, then the TPC track is matched to the CRT and selected~\cite{MicroBooNE:2019lta}.

\begin{figure}[h!]
    \centering
    \begin{subfigure}{0.50\textwidth}
        \centering
        \includegraphics[width=\linewidth]{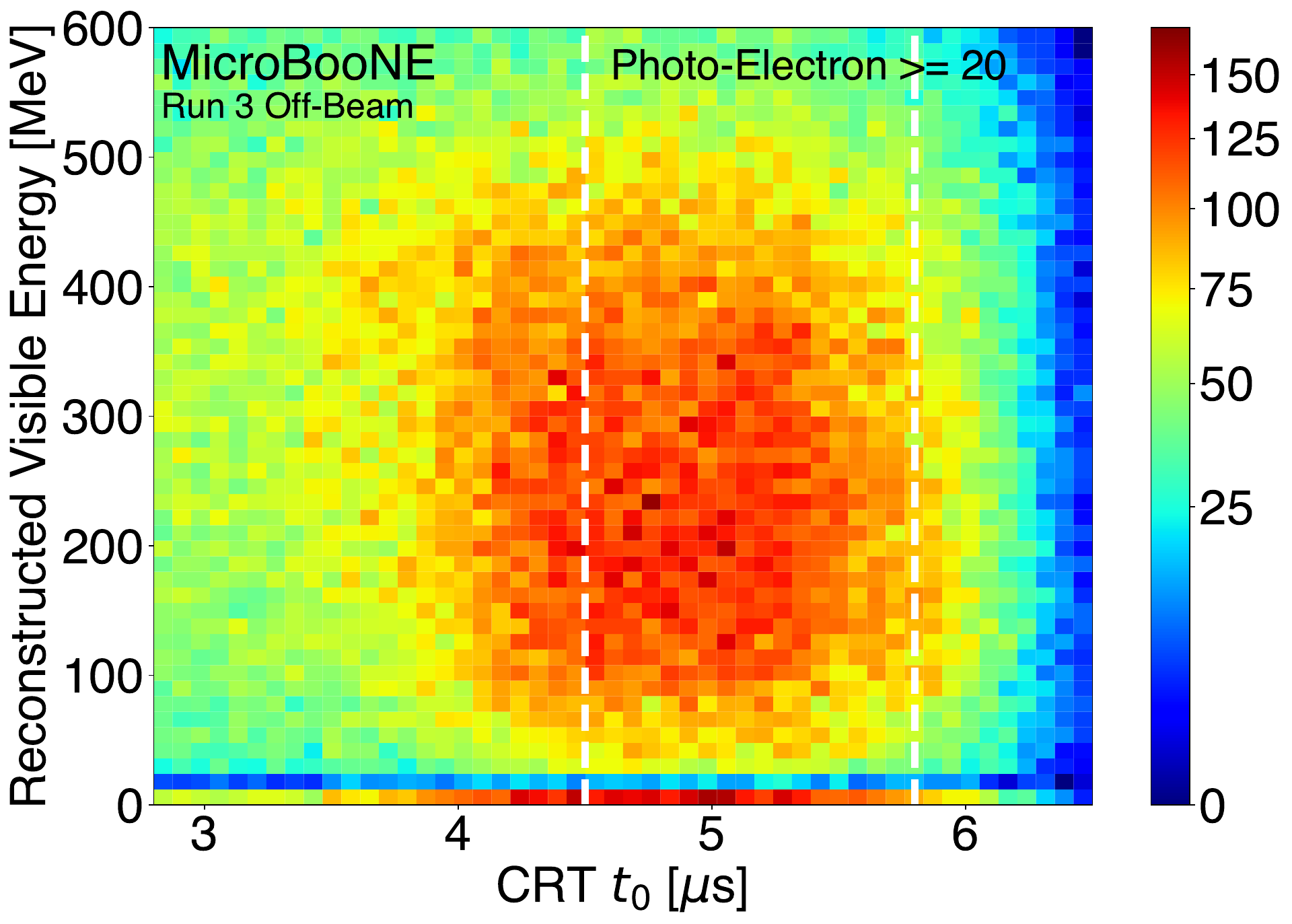}
        \caption{}
        \label{fig:ccrtt0_data_E_a}
    \end{subfigure}
    \hfill
    \begin{subfigure}{0.49\textwidth}
        \centering
        \includegraphics[width=\linewidth]{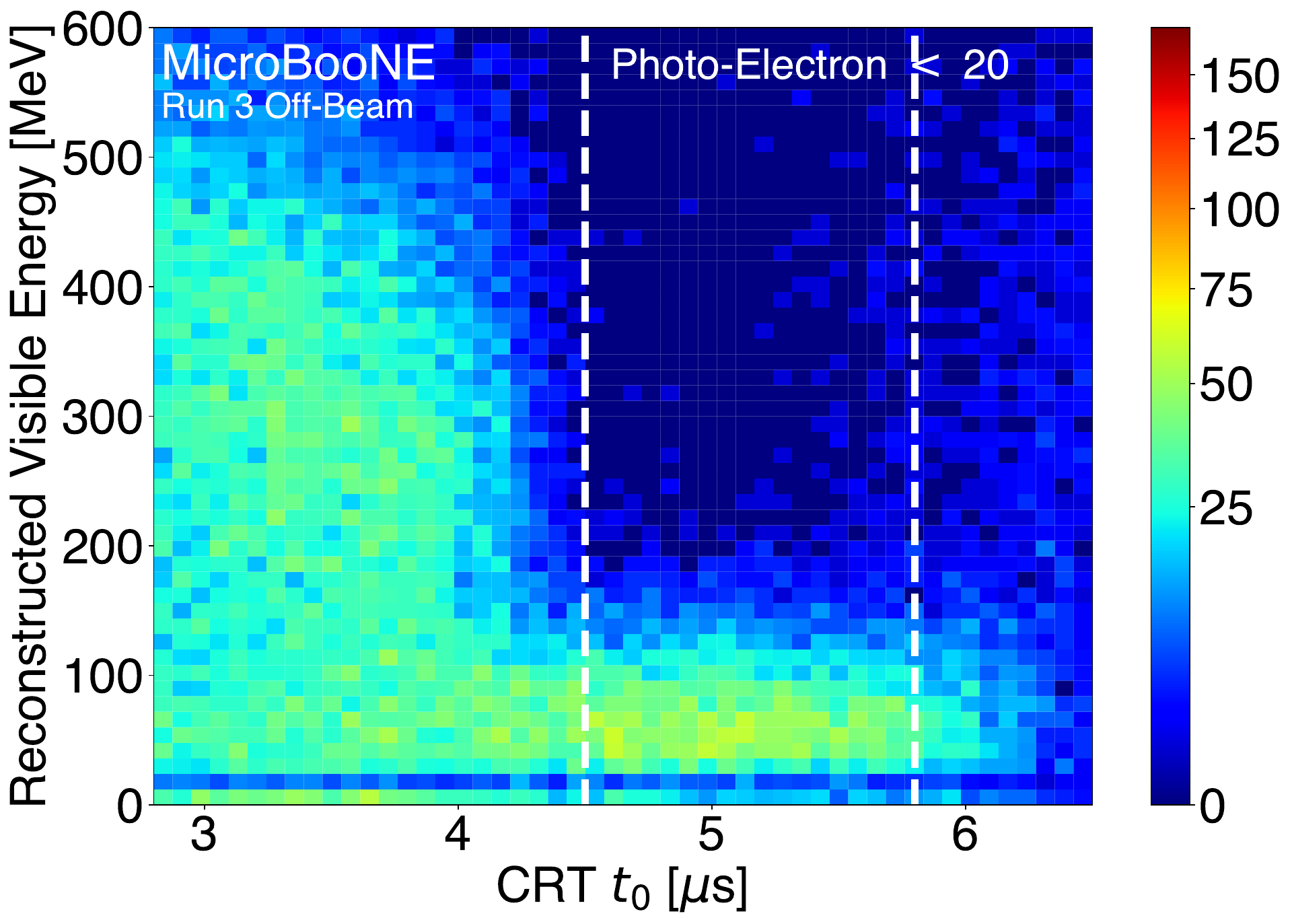}
        \caption{}
        \label{fig:ccrtt0_data_E_b}
    \end{subfigure}
    \caption{(a) Distribution of reconstructed visible energy for tracks with at least 20~PE and (b) for tracks with fewer than 20~PE, using the full data sample. In both cases, events with a CRT $t_0$ between 4.5 and 5.8~$\upmu$s are retained.}
    \label{fig:ccrtt0_data_E}
\end{figure}

Following this, a selection is applied to the CRT $t_0$ to select only events that produce light within the 1.6~$\upmu$s beam-window. This is done by examining the TPC-based measurement of the visible energy distributions as a function of CRT $t_0$ for events below and above 20~PE. As shown in figure~\ref{fig:ccrtt0_data_E}, events with $\geq 20$~PE lie within the desired energy range of 20--500~MeV and occur roughly between 4.4 and 6.0~$\upmu$s. In contrast, events with $< 20$~PE, in the same time range, exhibit energy below 100~MeV resulting in tracks further from the PMTs not depositing enough energy to exceed the light threshold. A significant population of $< 20$~PE events lies outside the 1.6~$\upmu$s trigger window and spans a broad energy range. As these events fall outside the acquisition window, their light is only partially recorded, and they are therefore excluded from this measurement. Similarly, out-of-time events near the outer edges of the dashed window are further restricted by tightening the selection to a 1.4~$\upmu$s window, retaining only events with CRT $t_0$ between 4.5 and 5.8~$\upmu$s. Finally, events below 20~MeV reconstructed energy arise from poorly reconstructed tracks traversing regions with unresponsive wires, and they are removed by subsequent selections.

\paragraph{Cathode piercing:}

\begin{figure}
	\centering
	\includegraphics[width=0.5\textwidth]{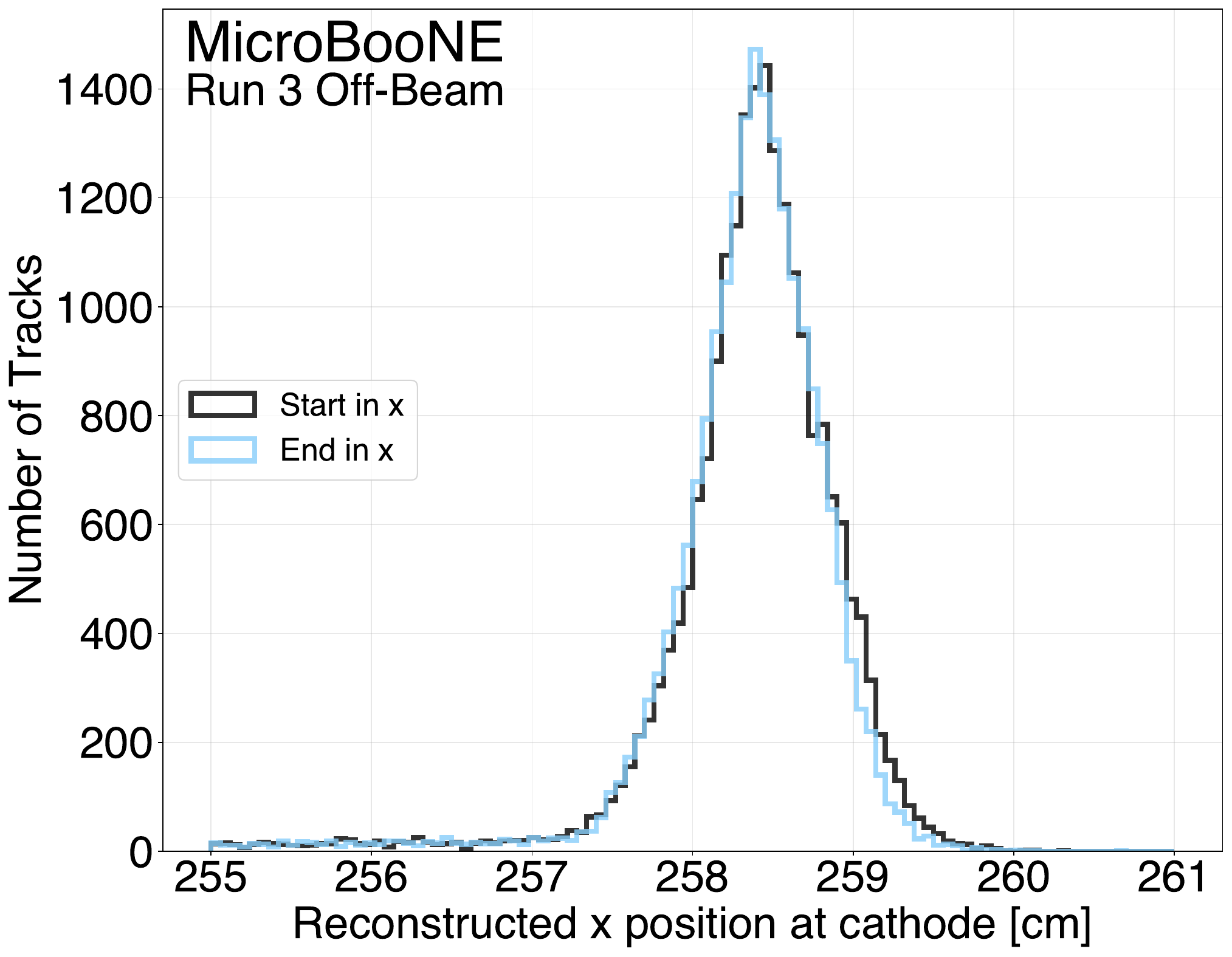}
	\caption{Distribution of the reconstructed start and end position in $x$ of the cosmic muon zoomed in at the cathode in data with both centred at the same value as expected.}\label{fig:x_pos_crt}
\end{figure}

Selected muons are required to enter or exit the TPC through the cathode to ensure piercing, minimally-ionising tracks. Events with reconstructed start or end positions beyond $x>257$ cm are selected centred at the location of the cathode, as shown in Figure \ref{fig:x_pos_crt}.

\paragraph{Track length and reconstructed charge:}

To ensure the reliability of the sample, this selection aims to remove tracks for which the TPC-reconstructed length or charge may be inaccurately measured, either due to limited track extent or detector effects. Short tracks less than 20~cm in length are generally unreliable and comprise a negligible fraction of the total sample, so they are not used in this measurement.

\begin{figure}
	\centering
	\includegraphics[width=0.65\textwidth]{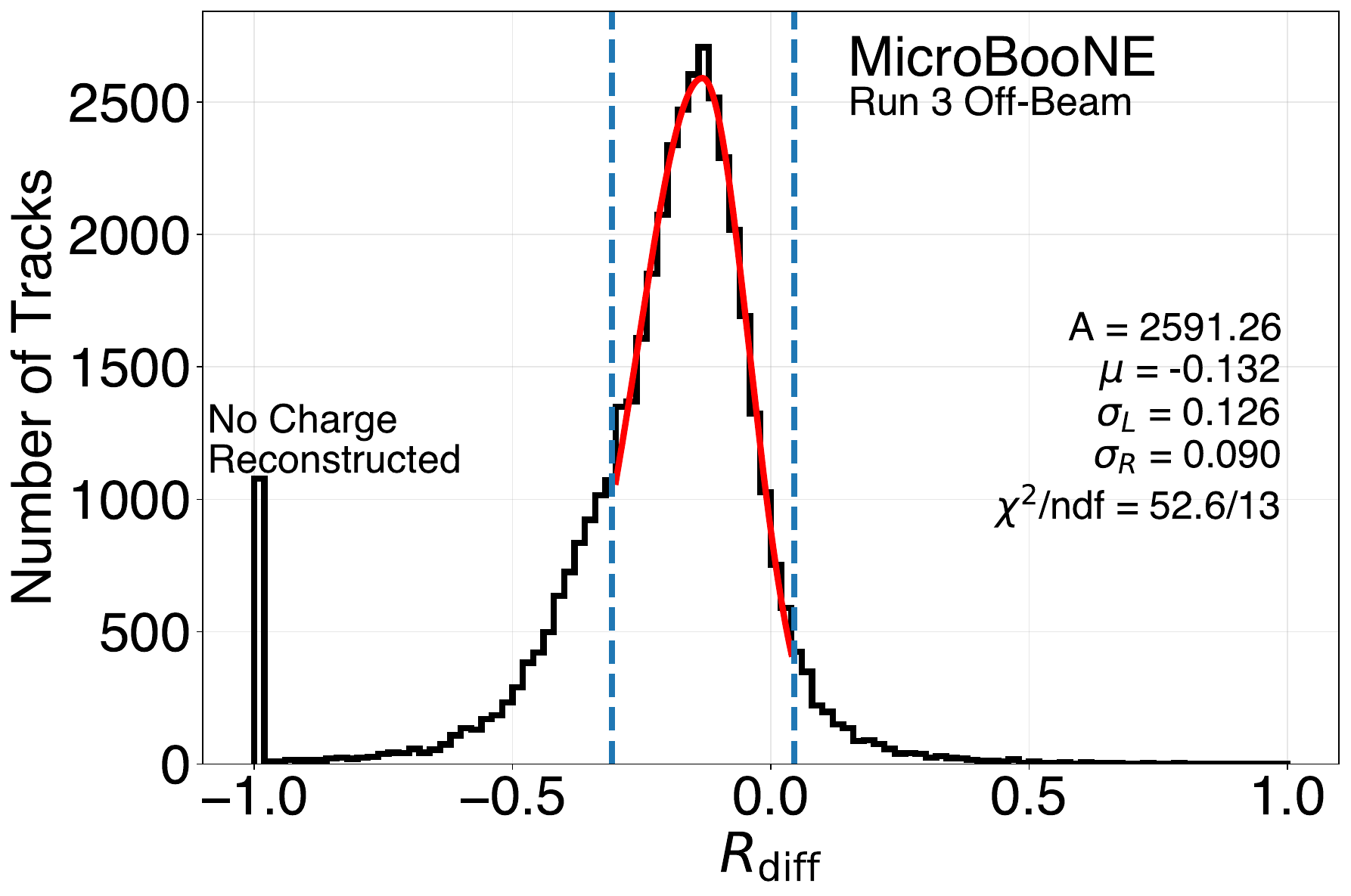}
	\caption{Distribution of the relative difference between $E_L$ and $E_Q$ with an asymmetric Gaussian fit used to define a 2$\sigma_R$ range (dashed blue lines) to exclude outliers, typically from tracks crossing unresponsive wire regions.} \label{fig:chargepl2}
\end{figure}

The track length of a particle traversing a LArTPC can be used to estimate its \textit{track energy} ($E_L$) via equation~\ref{eqn:energy_length}:

\begin{equation}
E_L = L\bigg\langle\frac{dE}{dx}\bigg\rangle\;,
\label{eqn:energy_length}
\end{equation}

\noindent where $L$ is the reconstructed track length, and $\big\langle\frac{dE}{dx}\big\rangle$ is the mean stopping power from the Bethe–Bloch eq.~\cite{ParticleDataGroup:2024cfk}. To obtain a more accurate measure of the deposited energy, which is directly correlated with the produced scintillation light, the energy is also calculated from the total charge collected on the TPC wires, $E_Q$, defined as:
\begin{equation}
    E_Q = QCWR\;,
\label{eqn:energy}
\end{equation}
where $Q$ is the measured charge in units of integrated wire waveform ADCs, $C$ is MicroBooNE’s calibration constant ($C=243$~\textit{e}/ADC~\cite{uboone_adc_e}), $W$ is the ionisation work function ($W=23.6$~eV/\textit{e}~\cite{Miyajima:1974zz}), and $R$ the is average recombination factor for muons ($R=0.6015$ at MicroBooNE’s field strength~\cite{david_thesis}). Because unresponsive wire regions can cause long tracks to appear with artificially low $E_Q$, a selection based on the relative difference between $E_L$ and $E_Q$:
\begin{equation}
R_\text{diff} = \frac{E_{Q}-E_{L}}{E_{L}}\;,
\label{eq:energy_ratio}
\end{equation}
was applied. Figure~\ref{fig:chargepl2} shows the distribution of $R_{\text{diff}}$, where an asymmetric Gaussian fit defines a 2$\sigma$ acceptance region (right-side width $\sigma_R$), represented by the dashed vertical lines, to exclude low-$R_{\text{diff}}$ outliers that would bias the lower-energy region.

A summary of the effects of the selections on the number of events is shown in table~\ref{table:selection}. It lists the various applied selections with both relative and absolute efficiencies, with the latter being the fraction of events passing each selection, and the former applied sequentially. The difference in relative efficiencies between data and simulation at the CRT–TPC matching stage is due to the simulated sample being artificially enriched with cathode-crossing tracks to improve matching efficiency, thereby reducing the computational resources required for event processing.

\begin{table}[ht]
    \caption{Overview of selections and efficiencies obtained. The relative efficiencies are calculated relative to the number of events at the previous stage of the selection while the absolute is relative to the total sample.} 
	\renewcommand{\arraystretch}{1.5}
	\centering 
	\begin{tabular}{c c c c c c c} 
		\toprule
		& \multicolumn{3}{c}{\textbf{Data}} & \multicolumn{3}{c}{\textbf{Simulation}}\\
		\cmidrule(lr){2-4}
		\cmidrule(lr){5-7}
		\textbf{\makecell{Selection}}&\textbf{\makecell{Number\\of Events}}& \textbf{\makecell{Relative\\Efficiency}} & \textbf{\makecell{Absolute\\Efficiency}}&\textbf{\makecell{Number\\of Events}} & \textbf{\makecell{Relative\\Efficiency}} & \textbf{\makecell{Absolute\\Efficiency}} \\ [0.5ex]
		\midrule
		\makecell{Total\\Sample}&376,183 &100\%  & 100\% &135,230 &100\% & 100\%  \\ [2ex]
		\makecell{CRT-TPC\\Matching}&81,135 &22\% & 22\% &93,345&69\% & 69\% \\ [2ex]
		\makecell{Cathode\\Piercing}& 40,939& 50\%& 11\% &51,517 & 55\%& 38\% \\ [2ex]
		\makecell{Track Length\\$> 20$~cm }&39,845& 97\%& 10.6\% & 51,510 &$>$99\% & 38\% \\ [2ex]
		\makecell{Charge\\Reconstructed} &31,468&79\% & 8.4\% & 40,452&79\% & 30\% \\ [1ex]
		\bottomrule
	\end{tabular}

	\label{table:selection}
\end{table}

\subsection{Method and results}

\begin{figure}[h!]
    \centering
    \begin{subfigure}{0.49\textwidth}
        \centering
        \includegraphics[width=\linewidth]{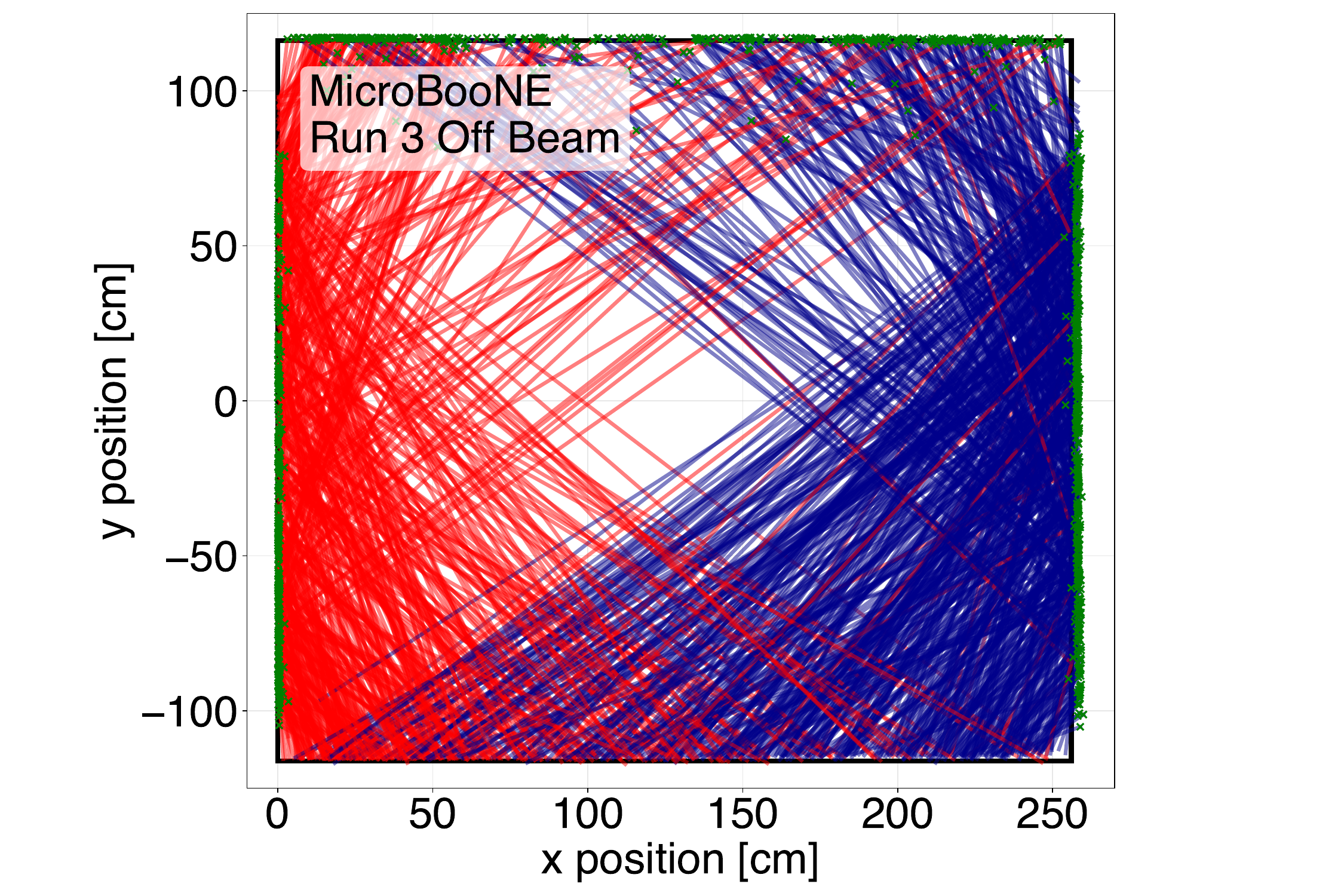}
        \caption{}
        \label{fig:graph_track_a}
    \end{subfigure}
    \hfill
    \begin{subfigure}{0.49\textwidth}
        \centering
        \includegraphics[width=\linewidth]{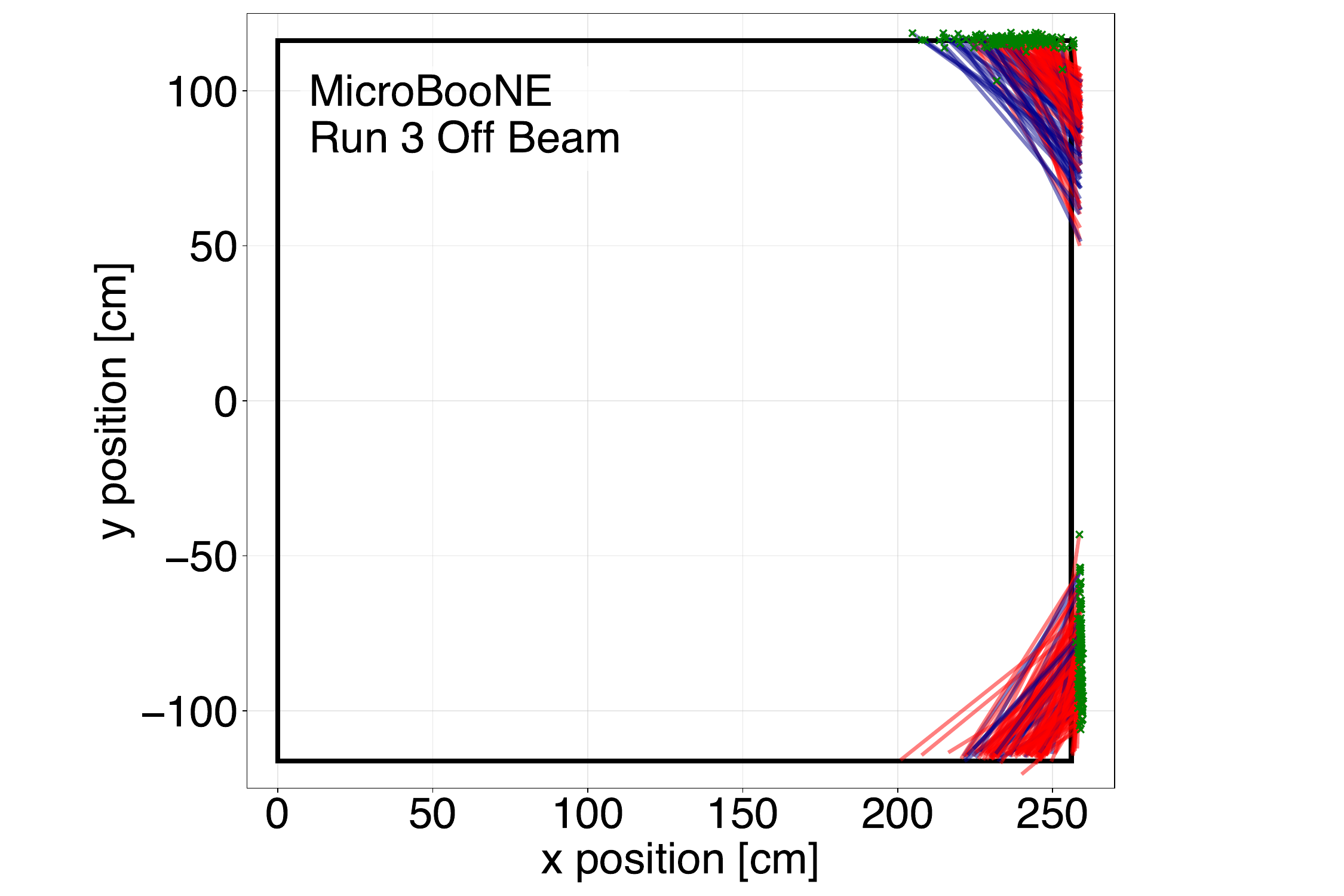}
        \caption{}
        \label{fig:graph_track_b}
    \end{subfigure}
    \caption{(a) Reconstructed start and end positions of 2,500 tracks in the YX plane. Red lines indicate tracks failing at least one selection, and blue lines are tracks passing the selection, and the green-$\times$ marks the reconstructed start position of all tracks shown. (b) Same distribution for a sample of 320 tracks with reconstructed energies below 200~MeV, highlighting the prevalence of short tracks near the cathode; red lines correspond to tracks failing the 20~PE threshold, while blue lines indicate those passing it.}
    \label{fig:graph_track}
\end{figure}

Figure \ref{fig:graph_track} summarises the reconstructed track positions in the YX plane. The selection effectively isolates tracks that traverse the cathode, including both long anode-to-cathode tracks and shorter ones near the top-right corner exceeding 20 cm in length. Most rejected tracks lie on the anode side or fail timing or charge requirements. The right panel focuses on a sample of sub-200 MeV tracks, illustrating that they cluster near the detector corners, consistent with the expected light yield distribution for these energies.

The scintillation light triggering efficiency is defined as the ratio between the number of selected tracks with energy in a bin $E_{j}$ that have more than 20~PE, and the total number of reconstructed tracks in that energy bin.

\begin{equation}
\textrm{Efficiency}[E_{j}] = \frac{\textrm{Tracks Producing $>$ 20 PE} [E_{j}]}{\textrm{All Tracks} [E_{j}]}.\
\label{eq:trig_eff}
\end{equation}

\begin{figure}
	\centering
	\includegraphics[width=0.65\textwidth]{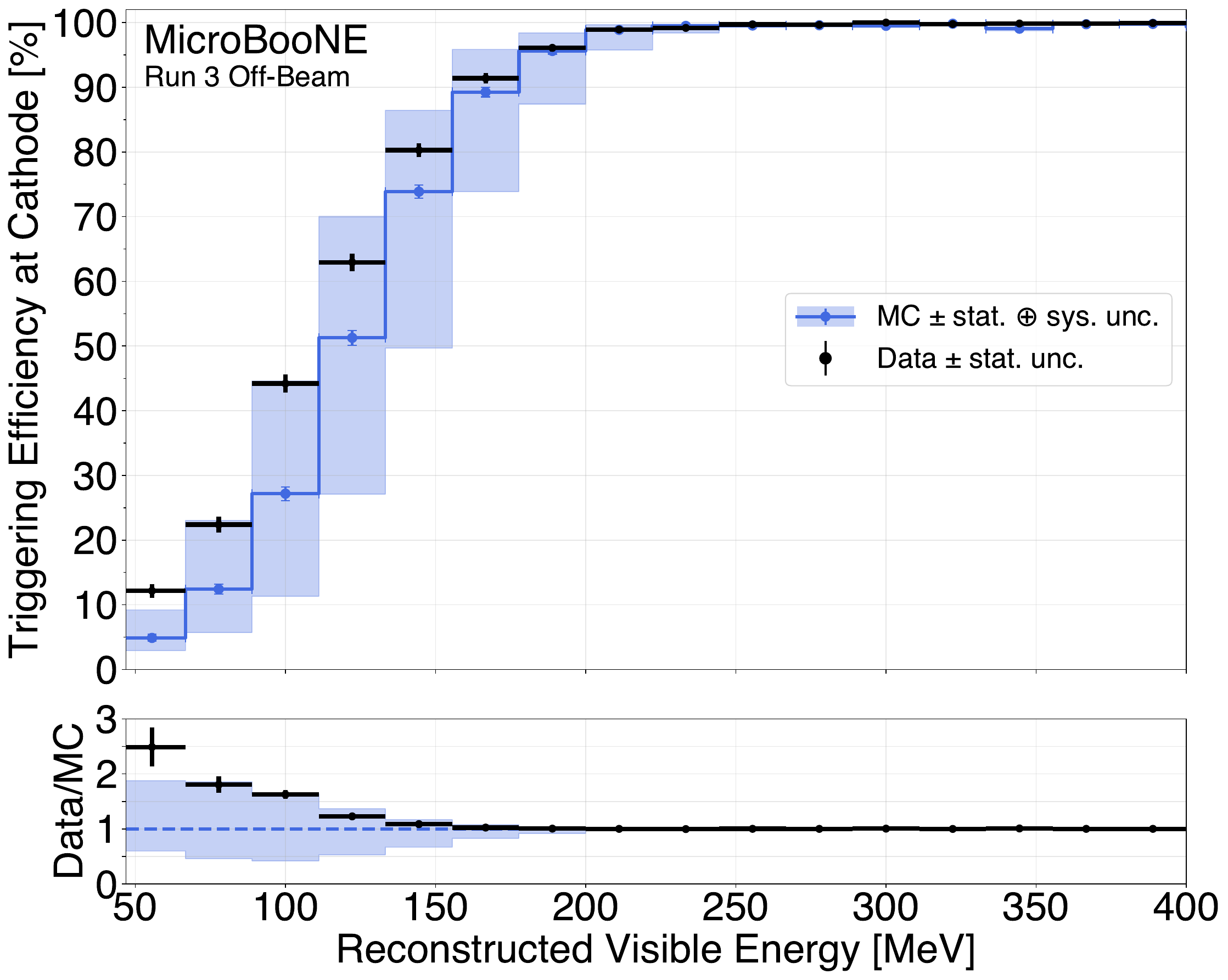}
	\caption{Top panel: scintillation light triggering efficiencies at the cathode using a 20~PE threshold, comparing data (black points with statistical uncertainties) to simulation (blue markers with statistical uncertainties bars) with the shaded region (band) indicates the total uncertainty. The sample represents the lowest light yield region in the detector furthest from the PMTs (cathode), where the trigger efficiency is lowest. Bottom panel: ratio of Data/MC with propagated uncertainties highlighting the overall agreement within uncertainties.}\label{fig:tot_eff_sys}
\end{figure}

\noindent The calculated scintillation light triggering efficiency at the cathode from 30~MeV to 400~MeV, using the charge-based energy estimator, using muons piercing the cathode, is shown in figure~\ref{fig:tot_eff_sys} for both the data and simulated events. The error bars indicate statistical uncertainties while the error band shows the total, statistical and systematic, uncertainty applied to the reconstructed number of PEs in simulation. For muon-like tracks of around 150~MeV of reconstructed visible energy, the efficiency is measured to be greater than 80\% and increases to nearly 100\% at 200~MeV. This energy range is particularly important for MicroBooNE’s flagship search for a low-energy excess in the few-hundred-MeV region. Even for events originating at the furthest distance from the PMTs, the system maintains nearly 100\% triggering efficiency, ensuring robust sensitivity in the region of greatest interest. The systematic uncertainty was obtained by looking at an orthogonal sample of point-like protons by comparing the PE/cm value of these events across the drift region. This study resulted in a 22\% difference between the data and the simulated PE/cm parameter, which describes the level of light mismodelling of the simulation in MicroBooNE.

The asymmetry in systematic uncertainty arises from applying the 22\% systematic directly to the number of PEs before comparison with the 20~PE threshold. Depending on the initial PE count of the simulated track, increasing or decreasing it by 22\% may not shift the value across the threshold. The largest uncertainty is observed near 125~MeV, where the rise in efficiency is sharpest, effectively marking an inflection point. In this region, the mean of the PE distribution lies close to the 20~PE threshold, so small fluctuations have a large impact: modest changes in PE shift some events from below to above the threshold (and vice-versa), effectively migrating events between selected and rejected samples once the uncertainty is applied. At energies above and below this range, the mean PE values are sufficiently far from the 20~PE threshold, minimizing migration effects and thereby reducing the overall uncertainty.

In this study, the scintillation-light triggering efficiency was evaluated in the lowest light-yield regime, accounting for a 50\% overall light-yield reduction for events occurring near the cathode. Even under these limiting conditions, the triggering efficiency remains high for energies above 125~MeV. These results, consistent with simulated efficiencies shown in figure~\ref{fig:sim_trig}, which predict a near 100\% efficiency at 20~PE across most of the active volume for energies above 50~MeV, confirm that the trigger system robustly captures the energy range most relevant to MicroBooNE’s flagship analyses. It is important to note that the sample of events chosen here unavoidably generates light outside of the TPC. The impact of this was evaluated by extending the tracks outside of the TPC to the cryostat boundaries where the light from these regions is expected to only contribute around 10\% of the total amount of light, smaller than the uncertainty due to mismodelling.

\begin{figure}
	\centering
	\includegraphics[width=0.65\textwidth]{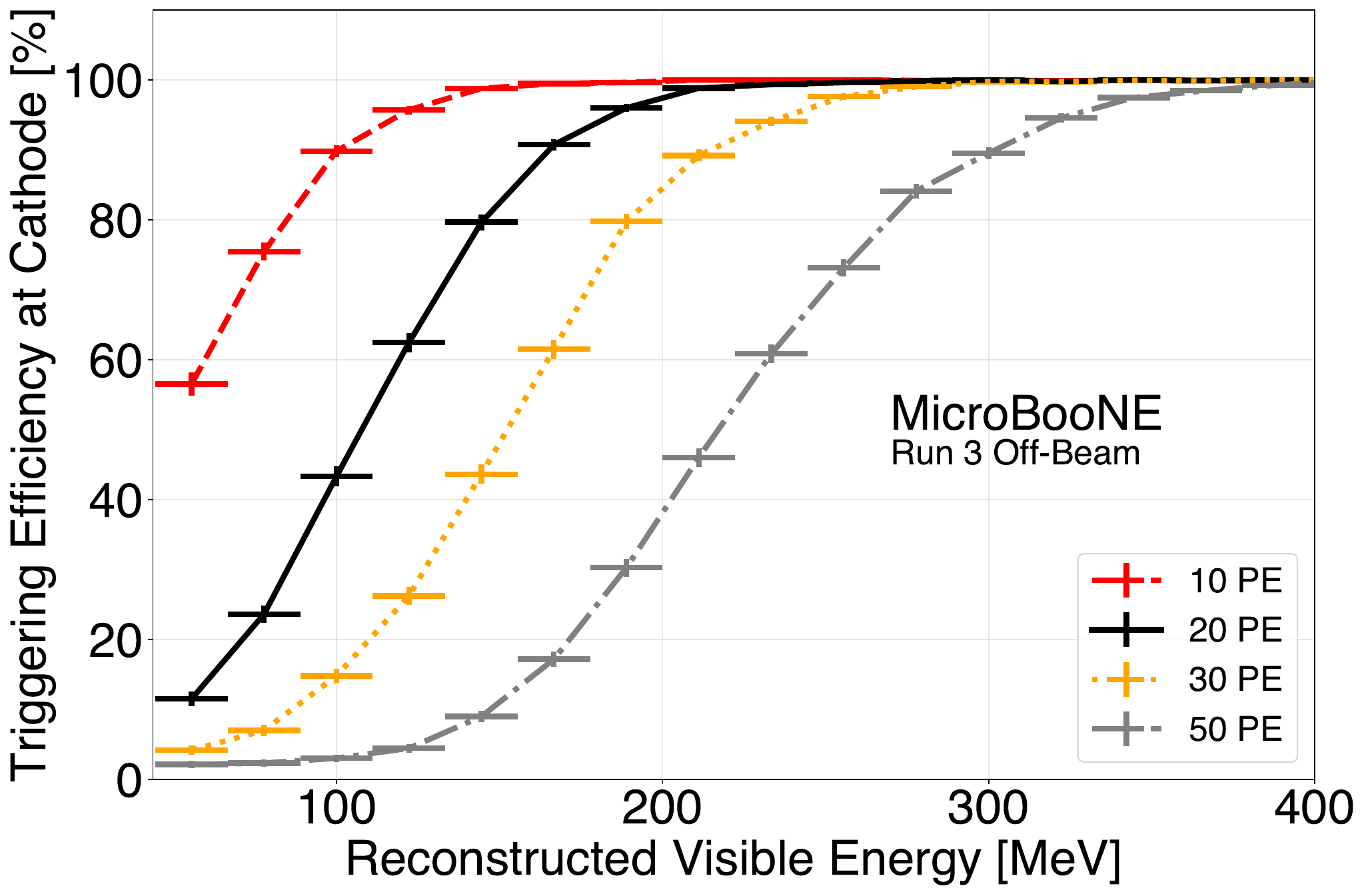}
	\caption{Scintillation light triggering efficiency at the cathode for PE thresholds from 10~PE to 50~PE. The 20~PE threshold corresponds to the offline trigger used in MicroBooNE analyses. Events down to the $\sim$5~PE online threshold are recorded.}
\label{fig:tot_eff_sys_diff_pe}
\end{figure}

To further understand the effect of the chosen PE threshold, the data triggering efficiency was recalculated for three other PE threshold values at 10~PE, 30~PE, and 50~PE as shown in figure~\ref{fig:tot_eff_sys_diff_pe}. The main observation is that reducing the PE threshold to 10~PE increases the efficiency of lower-energy events in this region of low light in MicroBooNE. This curve represents a conservative lower bound, with actual detector efficiency expected to be higher at these energies throughout the full detector as supported by figure~\ref{fig:sim_trig}, where the simulated efficiency is already high at 20~PE in most of the detector.

\section{Light response systematic uncertainties}

The recorded events used in MicroBooNE's $\mathcal{O}$(100~MeV) scale physics analyses are selected by the scintillation light trigger and as a result are impacted by changes in the amount of light observed. To account for uncertainties in the detector's optical response, dedicated light response detector systematic variations are used. MicroBooNE considers three distinct light-specific detector systematics: a uniform light yield decrease; an alternative Rayleigh scattering length; and a decreased attenuation length (Run 3 and later only). For each light response systematic, variation samples are produced making use of alternative photon lookup libraries. These are then passed through the remainder of the standard simulation and reconstruction to allow the impact of the variation at analysis level to be assessed. The light response detector systematics are designed to be complementary to the charge response detector systematics described in detail in ref.~\cite{MicroBooNE:2021roa}.

The first light response variation considers a uniform light yield decrease due to an overall mismodelling of the light compared with the data in addition to the time-dependent effects that are accounted for by calibration. To assess the level of uncertainty from this, events producing light at the anode or at the cathode were compared in data and simulation. Ratios of simulation to data at different values of drift distance were calculated for the amount of photoelectrons measured per centimetre of track length. These ratios showed an average 25\% over-prediction in the simulation. To assess the uncertainty arising from this, a photon lookup library with a uniform 25\% decrease in the light yield is used.

\begin{figure}[h!]
 \centering
 \includegraphics[width=0.7\textwidth]{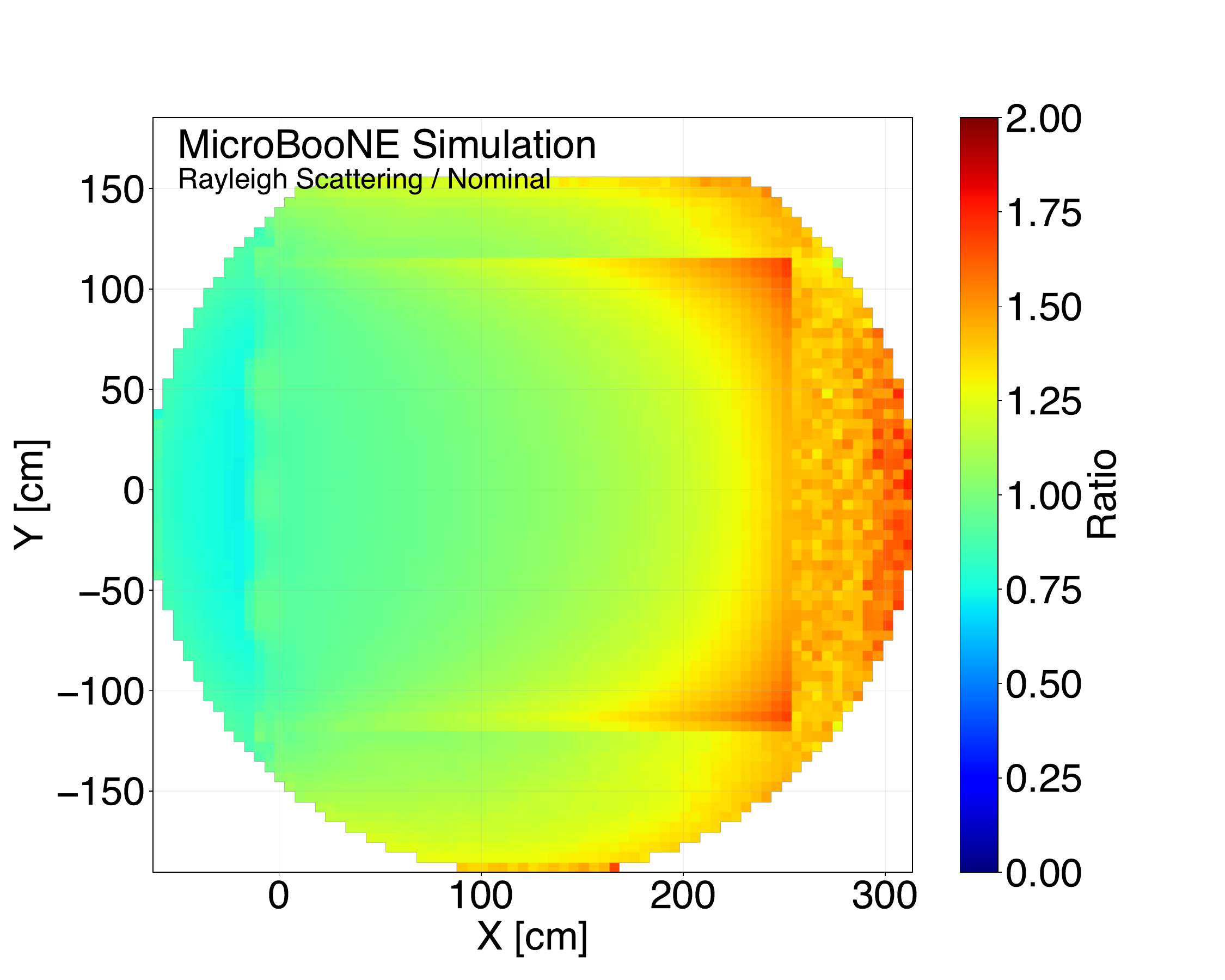}
 \caption{Sliced view in the XY plane of MicroBooNE detector's 120 cm Rayleigh scattering length photon visibility library variation used in the simulation of the photons.}
  \label{fig:RSL_library}
 \end{figure}

The second light response variation considers an alternative Rayleigh scattering length. MicroBooNE's simulation uses a Rayleigh scattering length of 66\,cm~\cite{Grace:2015yta}. More recent measurements suggest that the scattering length may be closer to 100\,cm~\cite{Babicz:2020den}. This would lead to position-dependent differences in the light observed compared with simulation. To assess the impact of this, the light yield variation due to different Rayleigh scattering lengths was studied using a MicroBooNE-like geometry. A simplified semi-analytical model~\cite{Garcia-Gamez:2020xrv} prediction was generated increasing the Rayleigh scattering length to 120\,cm. This was used to create an alternative light response map scaled by the ratio between a 120\,cm scattering length and the nominal 66\,cm to generate a much larger scattering length scenario closer to the currently accepted value of 100\,cm. Figure~\ref{fig:RSL_library} shows the ratio of this alternative photon lookup library to the nominal photon lookup library. The change of the Rayleigh scattering length results in a decrease in visibilities close to the PMTs ($x=0$~cm) and an increase farther from the PMTs. This has the largest impact close to the cathode ($x=256$~cm) and near the detector corners. The fluctuations seen in the ratio beyond the opaque cathode are as a result of the very small visibilities in this region, where small changes to extremely small visibilities result in larger changes to the ratio. This region where the light cannot, in most cases, reach the PMTs is disregarded.

\begin{figure}[h!]
 \centering
 \includegraphics[width=0.7\textwidth]{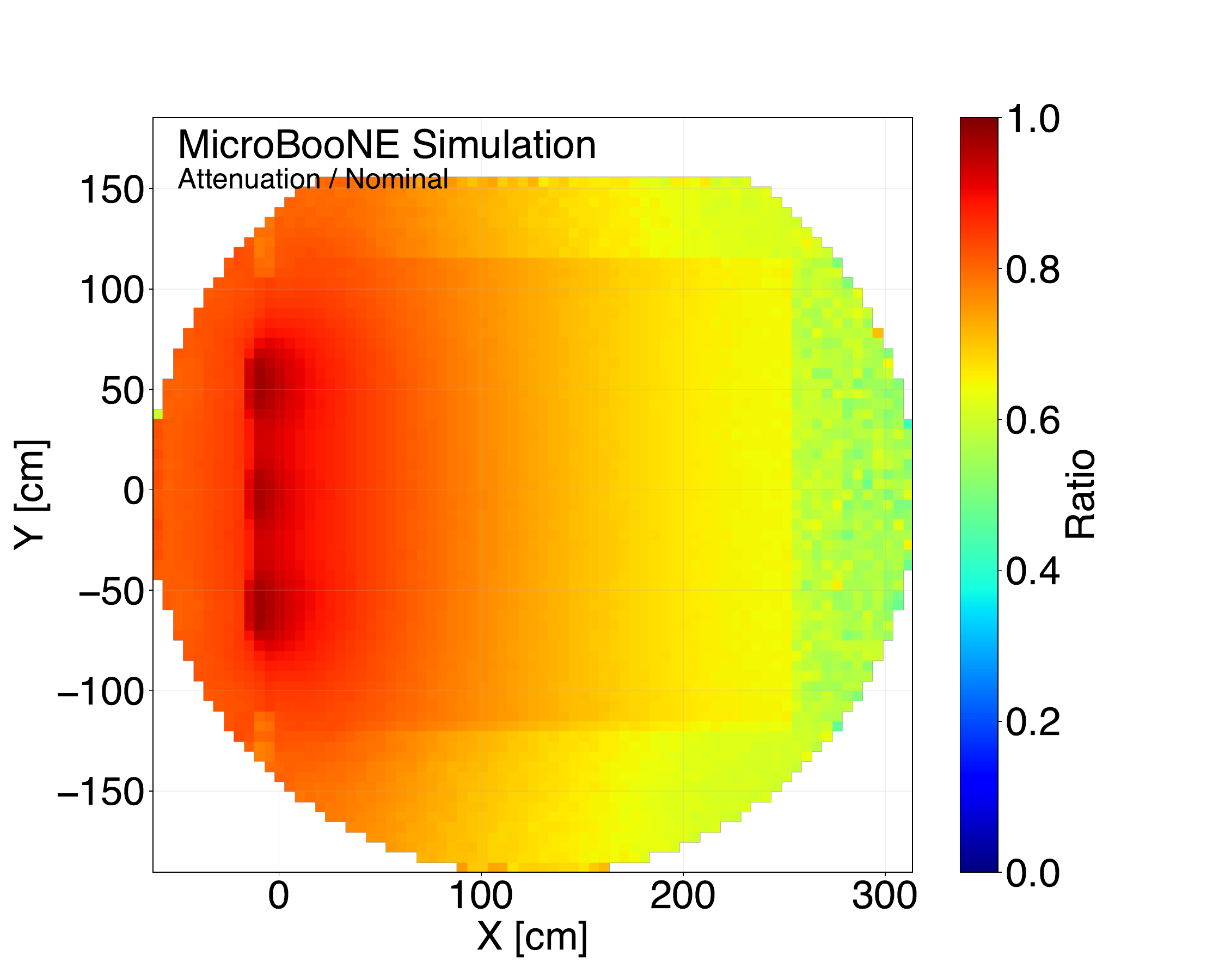}
 \caption{Sliced view in the XY plane of MicroBooNE detector's 8~m attenuation photon visibility library variation used in the simulation of the photons. The effect of the shorter attenuation length is seen by the reduction in the visibility in voxels especially at further distances.}
  \label{fig:attn_library}
 \end{figure}

Finally, the third light response variation considers a decreased attenuation length. The most significant light yield decline occurs between Run 1 and Run 3 of MicroBooNE's physics data collection. As discussed in section~\ref{sec:lydecline}, this is observed to have a dependence on the position of the interaction along the drift direction. This could correspond to a difference in attenuation length, potentially due to absorption as a result of the introduction of contaminants in the argon. Due to the uncertainty on the origin of this effect, it is not incorporated into the light yield decline calibration and therefore leads to a potential source of time-dependent systematic uncertainty. To evaluate the impact of this, an attenuation length variation is used. The variation was created by scaling the central-value light response map to have an absorption length of 8\,m compared with the original 20\,m. This additional systematic uncertainty is then applied only for samples from Run 3 onward. Figure~\ref{fig:attn_library} shows the ratio of this alternative photon lookup library and the nominal photon lookup library. The regions far from the PMTs have a decreased visibility probability due to the shorter attenuation length. As per the previous variations, the region beyond $x=256$~cm is disregarded due to very small visibilities.

\section{Discussion of light-related observations}{\label{sec:spe_rate}}

MicroBooNE has observed two puzzles regarding its scintillation light: a light yield decline concentrated in the first two years of running; and a high SPE rate $\mathcal{O}$(200~kHz) in each PMT that is also declining over time. The sources of these effects are currently unknown despite various investigations, some of which are currently ongoing. However, it is possible to assess their impact on the physics analyses performed by MicroBooNE and determine whether the included systematic uncertainties are sufficient. 

\subsection{Light yield decline}
\label{sec:ly_decline_discussion}

Exploration of the cause of the light yield decline is currently the subject of active investigation. One possible cause could be the slow introduction of impurities into the argon that would affect the light yield. However, this does not impact the ionisation where no time correlations have been observed between the electron lifetime and the light yield decline. As shown in figure~\ref{fig:light_response_change}, a slow introduction would have started from the start of operation in the fall of 2015 and accelerated in September 2016. The effect on the light yield then stabilised around early 2018. Impurities can both quench the slow component of the argon scintillation light, reducing the amount of late-light produced, and absorb the light during propagation to the PMTs, resulting in a position variation~\cite{Acciarri_2010, R_Acciarri_2010, B_J_P_Jones_2013_1, B_J_P_Jones_2013_2, CALVO2018186}. However, on-going studies suggest that we do not see a significant corresponding difference in the late-light time profile over this period that would be indicative of a contaminant being introduced. 

To further test the contaminants hypothesis, a sample of the MicroBooNE argon was collected in May 2021 and then analysed by the CIEMAT dark matter group in July 2021. A first analysis of these measurements was performed by colleagues at CIEMAT~\cite{Santorelli:2023vzj} using inductively coupled plasma mass spectrometry. This did not find any sufficiently large concentrations of contaminants in the detector that could explain the light response decline. A second sample of argon was taken in March 2023 and sent to Oneida Research Services, a commercial company expert in gas analysis. They performed both gas chromatography-mass spectroscopy~\cite{gcms} and internal vapor analysis~\cite{IVA} tests with the goal of looking for traces of organic and inorganic contaminants. Similarly, no sufficiently large concentration of contaminants was identified in this measurement. Further results on this topic will be presented in a dedicated future publication.

Another possible cause is degradation of either the PMTs or of the wavelength-shifting coating which covers the acrylic plates mounted in front of the PMTs. A possible indicator of this could be non-uniform changes in the light response decline in different regions of the detector or for different PMTs. To assess this, the ACPT samples described in section~\ref{sec:lydecline} were divided into five subsets along the detector length corresponding with the five PMT rosettes, illustrated in figure~\ref{fig:tpc_light}. The change in the light response over time was then measured for each subsample using only light seen in the PMTs in the corresponding rosette. No significant difference was observed in the light yield decline in any of the subsamples, suggesting the decline is uniform across the detector and across the PMTs. More fine-grained studies are difficult with the ACPT samples due to the extended energy depositions produced by the through-going cosmic-ray muons as well as the geometric constraints from the requirement to pass through the anode or cathode. On-going studies using alternative samples that produce more isolated energy depositions may provide further insights. These studies will be presented in a dedicated future publication.  

\subsection{Single photoelectron rate in MicroBooNE}
To measure the SPE rate of all PMTs across run periods, a stricter SPE pulse-finding algorithm based on the one presented in section~\ref{sec:pmt_gain} was implemented that uses the gain calibration to purify the SPE selection. A window in the PMT waveform is opened when a threshold of half the PE amplitude is crossed above the baseline. To reduce the spread of SPE ADC values, only pulses with amplitudes lower than 1.75~$\times$ the PE amplitude are recorded as SPE pulses. A dead-time of 469~ns (30 time ticks) is required following each pulse of any size in the waveform, to avoid selecting after-pulses as SPE candidates. The selection is finalised by removing waveforms with significant (>50~ADC) pulses above baseline and waveforms with baselines above or below the monthly PMT baseline trend by an amount exceeding five times the RMS noise.

The number of SPE pulses identified per waveform is recorded for each PMT over each month. A Poisson fit is applied to the monthly distribution to extract the mean number of pulses per waveform. The mean is used to approximate a correction for the dead time implemented by the algorithm and determine the effective search time per waveform. The SPE rate is then calculated as the mean number of pulses divided by the search time of the algorithm. Overall, an average SPE rate $\mathcal{O}$(200~kHz) was measured across all PMTs resulting in an integrated SPE rate of 8~MHz.

\begin{figure}
\centering
\includegraphics[width=0.7\textwidth]{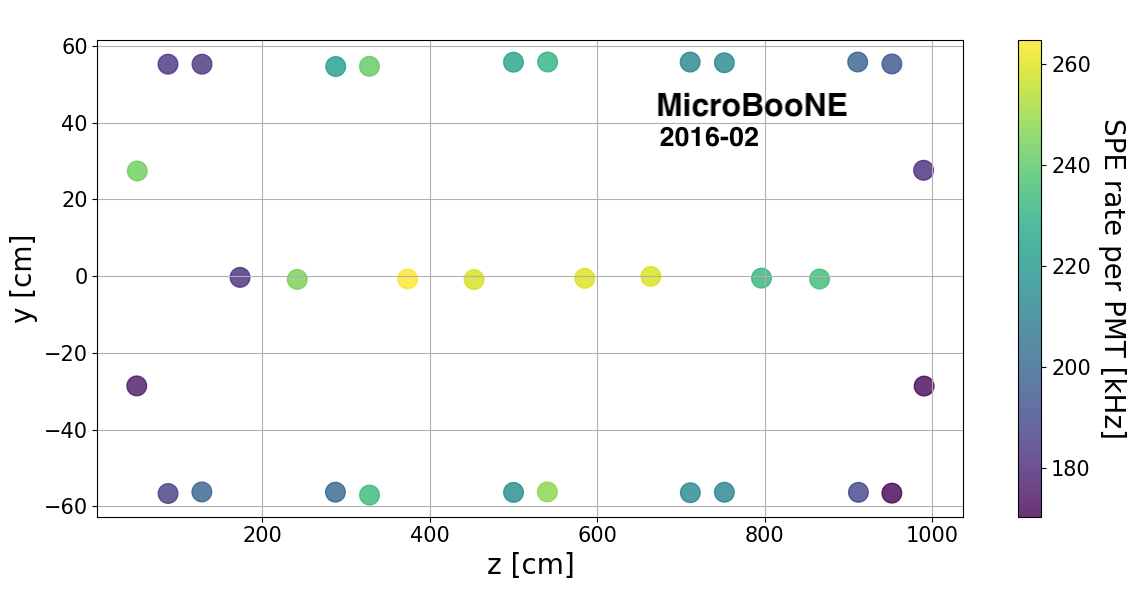}
\caption{Measured SPE rate for each MicroBooNE PMT in February 2016. A clear correlation is observed between the measured SPE rate and the PMT optical acceptance to liquid argon: PMTs positioned closer to the detector centre, with a larger active argon volume in view, exhibit higher SPE rates.}
\label{fig:spe_pmt_2D}
\end{figure}

As a representative example, a 2D map of the placement of each PMT in the detector and its respective measured SPE rate in February 2016 is shown in figure~\ref{fig:spe_pmt_2D}. PMTs closer to the middle of the detector have a higher rate of SPEs compared to those in the corners. This indicates a geometrical dependence on the  argon volume visible to each PMT.

From dedicated studies, a SPE rate of $\sim$50~kHz per PMT was expected in MicroBooNE based on standard radiological isotopes that are present in LArTPCs (e.g., argon-39, krypton-85, radon-222), as well as the expected intrinsic PMT noise. In addition to the higher measured SPE rate than expected  MicroBooNE has also observed two distinct interesting behaviours: a time-dependent decline throughout operation, and a correlation with the strength of the electric field. Both are discussed in this section.

\subsubsection{Single photoelectron rate measurement over time}

\begin{figure}
\centering
\includegraphics[width=0.7\textwidth]{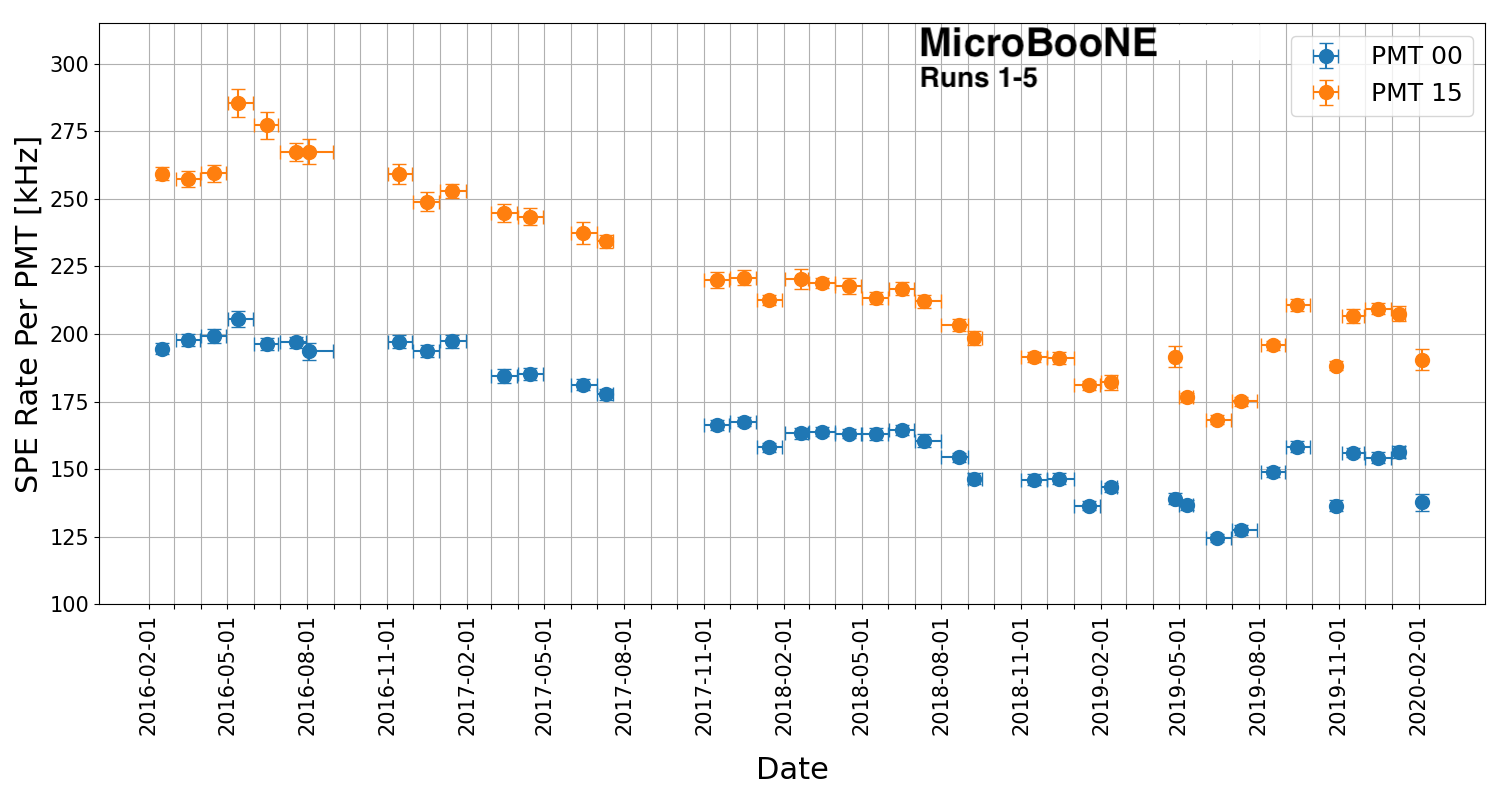}
\caption{Measured SPE rates for two MicroBooNE PMTs with the lowest and highest average rates over the full data-taking period. Both channels show a consistent decline with fluctuations correlated to variations in argon purity and cryogenic system filter regenerations.}
\label{fig:spe_pmt_00}
\end{figure}

\begin{figure}
\centering
\includegraphics[width=0.7\textwidth]{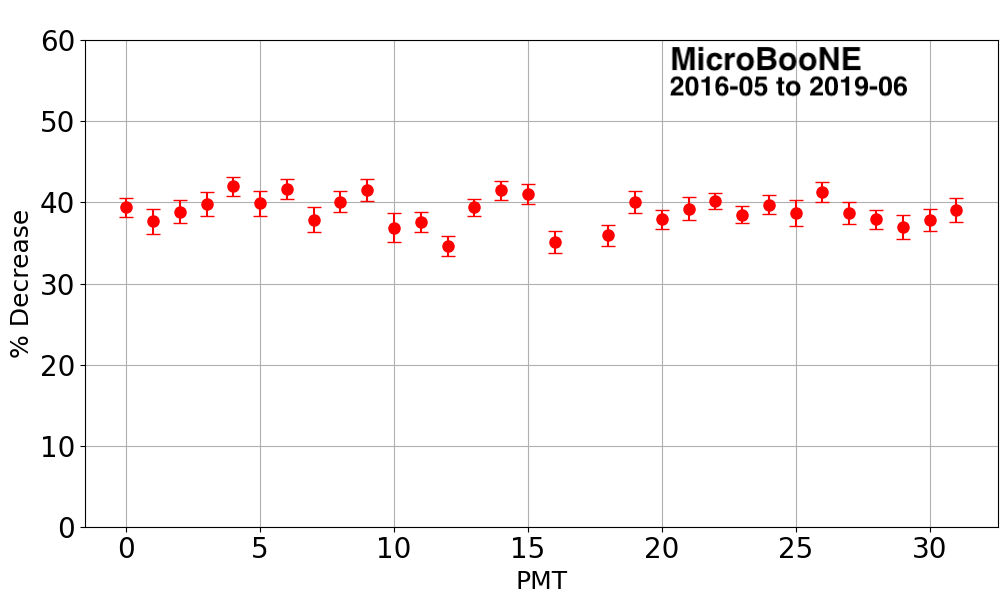}
\caption{Measured decline in SPE rates for all MicroBooNE PMTs between May 2016 and June 2019, showing an average decrease of approximately 40\% over three years.}
\label{fig:spe_pmt_decline}
\end{figure}

A measurement of the SPE rate throughout the full data taking period was made. Figure \ref{fig:spe_pmt_00} shows the measured SPE rate for two representative MicroBooNE PMTs, selected to illustrate channels with relatively low and high SPE activity with one in the corner and in the middle of the volume, respectively. The gaps in the time series correspond to periods of reduced liquid argon purity or to intervals when the PMTs were offline due to power outages or detector maintenance. For both of the example PMTs, the overall trend shows a gradual and stable decline in the measured rate over time. This behaviour is consistent across all PMTs, with an average reduction of approximately 40\% between May 2016 and June 2019, as shown in figure \ref{fig:spe_pmt_decline}. The SPE rate trend follows the variations in the measured light yield. The large gaps (e.g. September 2016, September 2017) coincide with low argon purity inferred from electron lifetime measurements where no high-statistics SPE measurements could be made while the upward trend in August 2019 corresponds to the filter regeneration mentioned in section~\ref{sec:lydecline}. If the source of the light response decline is the introduction of a contaminant into the detector, this could also correspondingly reduce the production of the SPE. These effects, observed simultaneously across the detector and for all PMTs, support the interpretation that the decrease in the SPE rate is possibly caused by changes in the liquid argon and overall detector conditions rather than solely progressive PMT ageing.

\subsubsection{Single photoelectron rate electric field dependence}

To understand the correlation between the measured SPE rate and the strength and polarity of the electric field through the TPC high voltage bias, dedicated R\&D runs at different electric fields were performed in summer 2021. As part of this, the negative (normal) polarity power supply was changed to a positive (reverse) polarity. This then allowed PMT data collection with a reversed electric field polarity. A commercial Glassman high voltage power supply (model PS/PK180P016XC5) with voltage and current maximum of +180~kV and 16~mA, respectively, was used. The output of the power supply was connected to a current-limiting resistor chain which was submerged in an aluminium container filled with transformer oil. This setup was done to serve as a low-pass filter, and a partition for the stored energy in the HV chain following what was previously done with the negative polarity power supply~\cite{MicroBooNE:2016pwy}. For safety of the detector and electronics, the current threshold of the power supply was set at 40~$\upmu$A.

Access to both positive and negative high-voltage supplies enabled a scan of the TPC drift voltage from -70~kV to +70~kV. To ensure safe operation during voltage ramp-up, a stepped procedure was adopted: 500~V increments up to 1~kV, 1~kV increments up to 5~kV, and 5~kV increments thereafter to the maximum voltage, applied for both polarities. TPC high voltage scans were performed between -70~kV and +70~kV at 10~kV intervals, with each voltage point maintained for a two-hour PMT data acquisition period. During this study, the wire bias was kept off during the reverse polarity scans for detector safety, and kept off during the normal polarity for symmetry of the measurement. This study represents the first operation of a LArTPC at positive polarity at such high voltage. 

\begin{figure}
\centering
\includegraphics[width=0.7\textwidth]{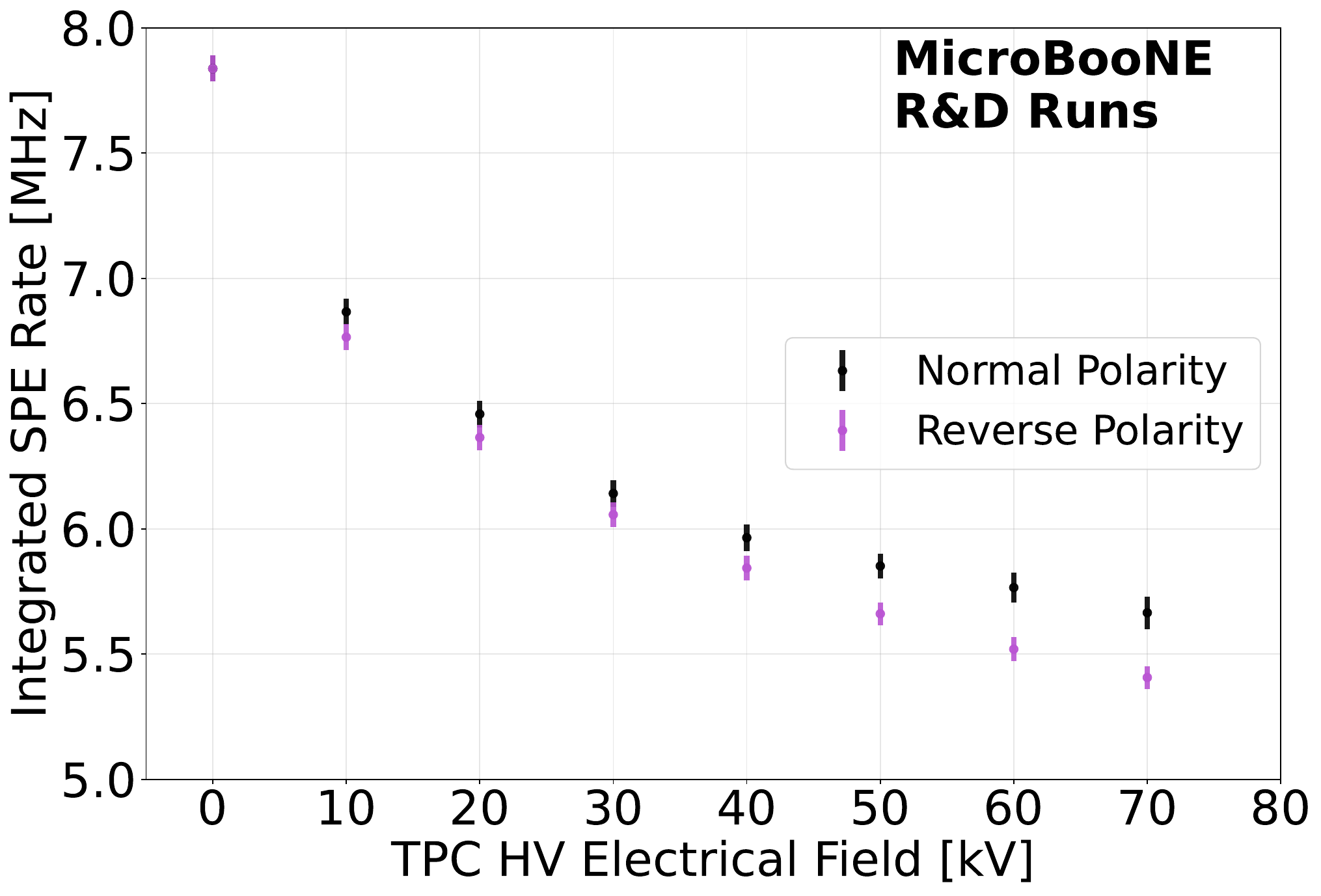}
\caption{Measurement of the MicroBooNE total integrated (all PMTs summed) SPE rate at different electric field value in both normal (negative) and reverse (positive) polarity of the HV. MicroBooNE's nominal HV is set at 70~kV, corresponding to an electrical field of 273~V/cm.}
\label{fig:spe_rate}
\end{figure}

The SPE rate summed over all PMTs is shown in figure~\ref{fig:spe_rate} with its dependence on the strength and polarity of the electric field. Typically, as the electric field strength grows, the measured SPE rate decreases due to reduced recombination and non-electron neutralisation. In normal polarity the ionisation electrons are drifted towards the anode, while in reverse they are drifted towards the cathode. In reverse polarity, the decrease in the SPE rate is larger than in normal polarity, underlining potential differences in physical processes. Analyses investigating modelling of the microphysics processes of scintillation light generation in liquid argon~\cite{Luo_2020} are currently ongoing as potential explanations. These explorations use a microphysics model which differentiates between the magnitude and polarity of the electric field to explain the high SPE rate. The differences observed in the two polarity modes will be the focus of a future publication.

\subsection{Impact on physics analyses}

The primary uses of scintillation light in MicroBooNE are to identify beam interactions in the detector and rejecting cosmic-ray backgrounds out-of-time with the beam. While selections on event samples vary, most analyses use a $\sim$20 PE prompt and $\sim$50 PE total light threshold on events. It is thus important to study whether the light yield decline has an impact on the number of neutrino candidates selected. The measurement of the scintillation light triggering efficiency from section~\ref{sec:trig_eff} demonstrated that the triggering efficiency remained high, even in the lowest-light region of the detector, after the decline in light yield had stabilised. In addition, the triggering efficiency matches the expectation from simulation within systematic uncertainties as shown in figure~\ref{fig:tot_eff_sys}. This result provides confidence that the triggering performance was consistently maintained at a high level throughout the detector’s operational lifetime. This is also evidenced by MicroBooNE's recently published analyses spanning the full five year dataset where no run-dependent effects due to the light yield decline were observed \cite{MicroBooNE:2024tym, MicroBooNE:2025pvb, MicroBooNE:2025phj}. Future analyses targeting lower-energy signals, such as searches for physics beyond the Standard Model, could lower the offline threshold to the $\sim$5~PE online threshold and gain the efficiency shown in figure~\ref{fig:trig_rate}.

Nonetheless, these studies and their findings further motivate careful monitoring of scintillation light-based detector performance variables during operations. These are important to consider for an experiment like SBND where the scintillation light plays an important role beyond simply triggering~\cite{SBND:2024vgn}, and for the future Deep Underground Neutrino Experiment (DUNE) both at the near- and far-detector complexes that are expected to run for decades.

\section{Summary}

In this article we have presented the long-term performance of scintillation light production and detection using five years of data from the MicroBooNE LArTPC detector. We describe the MicroBooNE light simulation, light-based triggering schemes, and light-based systematic uncertainties. We report two light-based calibrations using data from five years of operation: PMT gain calibration and time-dependent light yield stability calibration. A high $\mathcal{O}$(200~kHz) SPE rate is observed and reported here for the first time. This serves as an excellent calibration source for the PMT gains throughout the experiment’s lifetime, providing a stable and reliable source that did not require dedicated configurations during regular data collection campaigns. The PMT gain calibration was used to maintain an average amplitude gain of 20-23~ADC/PE after the first year of operation, with a spread of about 10\%. The light yield stability was measured across all run periods utilising samples of through-going cosmic-ray-induced muons. An approximately 50\% decline in the light yield was observed between late 2015 and late 2017. We report MicroBooNE's investigations into the origin of this decline, including measurements of two samples of argon for the presence of contaminants. The source of the decline remains unexplained. However, it is accounted for with a time-dependent calibration and has not impacted MicroBooNE's measurements. Next, we present an in-situ measurement of the scintillation light triggering efficiency down to 30\,MeV of visible energy. This measurement is performed in the lowest-light region of the detector furthest from the PMTs, and utilising data taken after the 50\% light yield decline. We find the triggering efficiency is high and stable within the energy range most relevant to MicroBooNE’s primary physics analyses. Finally, we characterise the observed high $\mathcal{O}$(200~kHz) SPE rate using data collected during dedicated R\&D runs. We observe an approximately 40\% decrease in the rate over time across all PMTs and we measure the dependence of the SPE rate on the TPC electric field strength and polarity. The results presented in this article provide an important benchmark of the long-term performance and stability of scintillation light systems in LArTPC neutrino detectors. This is essential for future LArTPC experiments expected to operate over many years including the Short-Baseline Neutrino Program~\cite{Machado:2019oxb} and the Deep Underground Neutrino Experiment~\cite{DUNE:2020lwj}. 

\acknowledgments
This document was prepared by the MicroBooNE collaboration using the resources of the Fermi National Accelerator Laboratory (Fermilab), a U.S. Department of Energy, Office of Science, Office of High Energy Physics HEP User Facility. Fermilab is managed by Fermi Forward Discovery Group, LLC, acting under Contract No. 89243024CSC000002. MicroBooNE is supported by the following: the U.S. Department of Energy, Office of Science, Offices of High Energy Physics and Nuclear Physics; the U.S. National Science Foundation; the Swiss National Science Foundation; the Science and Technology Facilities Council (STFC), part of United Kingdom Research and Innovation (UKRI); the Royal Society (United Kingdom); the UKRI Future Leaders Fellowship; the NSF AI Institute for Artificial Intelligence and Fundamental Interactions; and the European Union’s Horizon 2020 research and innovation programme under the Marie Sk\l{}odowska-Curie grant agreement No. 101003460 (PROBES). Additional support for the laser calibration system and cosmic ray tagger was provided by the Albert Einstein Center for Fundamental Physics, Bern, Switzerland. We also acknowledge the contributions of technical and scientific staff to the design, construction, and operation of the MicroBooNE detector as well as the contributions of past collaborators to the development of MicroBooNE analyses, without whom this work would not have been possible. 

\bibliographystyle{JHEP}
\bibliography{main}

\end{document}